\documentclass[11pt,a4paper,epsf,epsfig,psfrag]{article}
\usepackage{jheppub}
\usepackage{amsmath,graphicx,amsfonts, mathrsfs,amssymb}
\usepackage{amsmath,epsfig}
\usepackage{amssymb,amsfonts}
\usepackage{latexsym}
\usepackage{epsfig}
\usepackage{pdfsync}
\newbox\pippobox

\def\be{\begin{equation}}
\def\ee{\end{equation}}
\def\bea{\begin{eqnarray}}
\def\eea{\end{eqnarray}}

\def\lag{\langle}
\def\rag{\rangle}

\newcommand{\beq}{\begin{equation}}
\newcommand{\eeq}{\end{equation}}
\newcommand{\beqa}{\begin{eqnarray}}
\newcommand{\eeqa}{\end{eqnarray}}
\newcommand{\beqar}{\begin{eqnarray*}}
\newcommand{\eeqar}{\end{eqnarray*}}

\renewcommand{\eqref}[1]{(\ref{#1})}

\catcode`\@=12

\def\lag{\langle}
\def\rag{\rangle}

\newcommand\mr{\mathscr}

\title{Dynamical holographic QCD model for glueball and light meson spectra}
\author{Danning Li $^{a}$,  Mei Huang$^{a,b}$\\
$^{a}$ {Institute of High Energy Physics, Chinese Academy of
Sciences, Beijing, China} \\
$^{b}$ {Theoretical Physics Center for Science Facilities, Chinese
Academy of Sciences, Beijing, China} \\
}

\abstract{
In this work, we offer a dynamical soft-wall model to describe the gluodynamics and
chiral dynamics in one systematical framework. We firstly construct a quenched
dynamical holographic QCD (hQCD) model in the graviton-dilaton framework for the
pure gluon system, then develop a dynamical hQCD model for the two flavor system
in the graviton-dilaton-scalar framework by adding light flavors on the gluodynamical
background. For two forms of dilaton background field $\Phi=\mu_G^2z^2$ and $\Phi=\mu_G^2z^2\tanh(\mu_{G^2}^4z^2/\mu_G^2)$, the quadratic correction
to dilaton background field at infrared encodes important non-perturbative gluodynamics
and naturally induces a deformed warp factor of the metric. By self-consistently
solving the deformed metric induced by the dilaton background field, we find that
the scalar glueball spectra in the quenched dynamical model is in very well agreement
with lattice data. For two flavor system in the graviton-dilaton-scalar framework, the
deformed metric is self-consistently solved by considering both the chiral condensate
and nonperturbative gluodynamics in the vacuum, which are responsible for the chiral
symmetry breaking and linear confinement, respectively. It is found that the mixing
between the chiral condensate and gluon condensate is important to produce the correct light
flavor meson spectra. The pion form factor and the vector couplings are also investigated in the dynamical hQCD model. Besides, we give the criteria for the existence of linear quark potential from the metric structure, and show a negative quadratic dilaton background field is not favored
in the graviton-dilaton framework.}


\begin{document}
\maketitle

\section{Introduction}

The local quantum field theoretical description has made great success since it
was firstly developed in the quantization of electrodynamics and
further been developed and applied to the descriptions of elementary particles.
Nowadays, quantum chromodynamics (QCD) is accepted as the fundamental theory of
the strong interaction. In the ultraviolet (UV) or weak coupling regime of
QCD, the perturbative calculations agree well with experiment. However, in the
infrared (IR) regime, the description of QCD vacuum as well as hadron properties
and processes in terms of quark and gluon still remains as outstanding challenge
in the formulation of QCD as a local quantum field theory. In order to derive the
low-energy hadron physics and understand the deep-infrared sector of QCD from first
principle, various non-perturbative methods have been employed, in particular
lattice QCD \cite{Kogut:1982ds,Gupta:1997nd,Fodor:2012gf,Bloch:2003sk}, Dyson-Schwinger
equations (DSEs)\cite{Alkofer:2000wg,Bashir:2012fs}, and functional renormalization
group equations (FRGs)\cite{Wetterich:1992yh,Pawlowski:2005xe,Gies:2006wv}.

In recent decades, an entirely new method based on the anti-de Sitter/conformal
field theory (AdS/CFT) correspondence and the conjecture of the gravity/gauge duality \cite{Maldacena:1997re,Gubser:1998bc,Witten:1998qj} provides a revolutionary method
to tackle the problem of strongly coupled gauge theories.
Though the original discovery of holographic duality requires supersymmetry and
conformality, the holographic duality has been widely used in investigating hadron
physics, strongly coupled quark gluon plasma and condensed matter.
It is widely believed that the duality between the quantum field theory and quantum gravity
is an unproven but true fact. In general, holography relates quantum field theory (QFT) in
d-dimensions to quantum gravity in (d + 1)-dimensions, with the gravitational description
becoming classical when the QFT is strongly-coupled. The extra dimension
can be interpreted as an energy scale or renormalization group (RG) flow in the QFT
\cite{deBoer:1999xf,Skenderis:2002wp,deBoer:2000cz,Li:2000ec,Heemskerk:2010hk,
Faulkner:2010jy,Balasubramanian:2012hb,Adams:2012th}.

Many efforts have been invested in applying holography duality to describe the real QCD
world, e.g. for mesons \cite{EKSS2005,Karch:2006pv,mesons}, baryons \cite{baryons} and
glueballs \cite{glueballs}, see Refs. \cite{topdown-Reviews,review}
for reviews. It is well-known that
the QCD vacuum is characterized by spontaneous chiral symmetry breaking and color charge
confinement. The chiral symmetry breaking can be read from the mass difference between
the chiral partners of the hadron spectra, and the spontaneous chiral symmetry breaking
is well understood by the dimension-3 quark condensate $\lag\bar{q}q\rag$ \cite{NJL}
in the vacuum. The understanding of confinement remains a challenge \cite{Greensite}.
From the hadron spectra, confinement can be read from the Regge trajectories of hadrons
\cite{Regge,pdg}, which suggests that the color charge can form the string-like structure
inside hadrons. Confinement can also manifest itself by the linear potential between two
quarks at large distances \cite{Cornell}.

A successful holographic QCD model should describe chiral symmetry breaking,
and at the same time should describe both the Regge trajectories of hadron
spectra and linear quark potential, two aspects in the manifestation of
color confinement. Nonetheless, these important nonperturbative
features haven't been successfully accommodated in a unique hQCD model.

The current achievements of AdS/QCD models for hadron spectra are the hard-wall AdS/QCD
model \cite{EKSS2005} and the soft-wall AdS/QCD or KKSS model \cite{Karch:2006pv}.
In the hard-wall model \cite{Karch:2006pv}, the chiral symmetry breaking can be realized
by chiral condensation in the vacuum, however, the resulting mass spectra for the excited
mesons behave as $m_n^2\sim n^2$, which is different from the linear Regge behavior
$m_n^2\sim n$. In order to generate the linear Regge behavior, the authors of
Ref.\cite{Karch:2006pv} introduced a quadratic dilaton background, one can obtain the
desired mass spectra for the excited vector mesons, while
the chiral symmetry breaking phenomenon cannot be consistently realized
\cite{Huang:2007fv}.

Interesting progress was made in Refs. \cite{Colangelo:2008us, Gherghetta-Kapusta-Kelley},
where a quartic interaction term in the bulk scalar potential was introduced to incorporate
linear trajectories and chiral symmetry breaking. However, such a term was shown
\cite{Gherghetta-Kapusta-Kelley} to result in a negative mass square for the lowest lying scalar meson state,
which might cause an instability of the background.
In Ref.\cite{YLWu}, a deformed warp factor is introduced, which can cure the instability and maintain the linear behavior of the spectra.

With AdS$_5$ metric in the soft-wall model and its extended versions \cite{Colangelo:2008us, Gherghetta-Kapusta-Kelley,YLWu,Batell:2008zm,Gherghetta-Kapusta-Kelley-2,Afonin:2010hn,dePaula:2008fp,modified-dc} (except \cite{YLWu}), only Coulomb potential between the two quarks can be produced \cite{Maldacena:1998im}.
On the other hand, the linear quark potential can be realized in the Andreev-Zakharov
model \cite{Andreev:2006ct}, where a positive quadratic correction in the deformed warp
factor of ${\rm AdS}_5$ geometry was introduced. The linear heavy quark potential can
also be obtained by introducing other deformed warp factors as in Refs.
\cite{Pirner:2009gr,He:2010ye}.
The positive quadratic correction in the deformed warp factor in some
sense behaves as a negative dilaton background in the 5D action, which
motivates the proposal of the negative dilaton soft-wall model
\cite{Zuo:2009dz,deTeramond:2009xk}. More discussions on the sign of
the dilaton correction can be found in \cite{KKSS-2,Schmidt-pn-dilaton}.

It is noticed that the quadratic correction, whether appears in
the 5D action or in the deformed warp factor, indeed plays an important role to
realize the linear confinement, though only partly. Since both the Regge trajectories
of hadron spectra and linear quark potential are two aspects in the manifestation of
color confinement, they should share the same dynamical origin and should be realized
in the same holographic QCD model.

In the soft-wall model and its improved versions, the dilaton field or the deformed warp
factor are introduced by hand. In Ref.\cite{Li:2012ay}, we have successfully described
the chiral symmetry breaking, the Regge trajectories of hadron spectra and linear
quark potential in the graviton-dilaton-scalar coupling framework, in which the metric,
the field(s) and the potential(s) of the field(s) are self-consistently determined by
field equations, and one can self-consistently solve out the other two with one input.
There are at least three different ways to deal with the system in the literature:
1) Input the form of the field(s) to solve the metric structure
and the potential(s) of the field(s) \cite{Csaki:2006ji,Li:2012ay};
2) Input the potential(s) of the field(s) to solve the metric and the field(s) \cite{Gubser:2008ny,GKN};
3) Input the form of the metric structure to solve the field(s) and the potential(s)
of the field(s) \cite{Li:2011hp}.

This work is an extension of Ref.\cite{Li:2012ay}, and we intend to establish the relation
between the QCD dynamics including at IR and its induced geometry.
The paper is organized as follows: In Sec. \ref{sec-glueball}, we establish a quenched
dynamical hQCD model in the graviton-dilaton framework to describe the pure gluon system, and
by selfconsistently solve the deformed warp factor induced by the dilaton field, we get the
scalar glueball spectra;  We introduce the meson spectra in the KKSS model and its improved
versions in Sec.\ref{sec-hadronspectra-KKSS}; We then
develop the graviton-dilaton-scalar coupling framework for two flavor system
and investigate the hadron spectra in Sec.\ref{sec-hadronspectra-dhQCD}, and we also
investigate pion form factors and vector couplings in Sec. \ref{sec-formfactors}. We give
discussion and summary in Sec.\ref{sec-summary}.

\section{Pure gluon system: quenched dynamical soft-wall holographic model}
\label{sec-glueball}

At the classical level, QCD is a scale invariant theory, which is broken by
quantum fluctuations. The pure gluon part of QCD Lagrangian in 4-dimension
is described by
\begin{equation}
\mr{L}_{G} = -\frac{1}{4}G_{\mu\nu}^a(x)G^{\mu\nu,a}(x), \label{eq:lg}
\end{equation}
with
\begin{equation}
G_{\mu\nu}^a(x) = \partial_{\mu}{A_{\nu}^a(x)}-\partial_{\nu}{A_{\mu}^a}(x)+gf^{abc}{A_{\mu}^b}(x){A_{\nu}^c}(x).
\label{eq:gmunu}
\end{equation}
Where $A_{\mu}(x)$ is the gluon field with $a=1,\cdots,8$ the color indices.

In the vacuum, the scale invariance is explicitly broken, and the relevant degrees
of freedom of QCD at infrared are still poorly understood. Varies of vacuum
condensates provide important information to understand the non-perturbative
dynamics of QCD. For example, the gauge invariant dimension-4 gluon condensate
$\lag g^2G^2\rag$ has been widely investigated in both QCD sum rules and lattice
calculations \cite{Shifman:1978by,Boyd:1996bx,Schafer:1996wv}, and the non-vanishing
value of the condensate does not signal the breaking of any symmetry directly, but
rather the non-perturbative dynamics of strongly interacting gluon fields. In last
decade, there have been growing interests in dimension-2 gluon condensates $\lag g^2
A^2\rag $ in SU$(N_c)$ gauge theory and its possible relation to confinement
\cite{Celenza:1986th,Lavelle:1988eg,Lavelle:1992yh,Gubarev:2000eu,Verschelde:2001ia,
Chetyrkin:1998yr,Gubarev:2000nz,Kondo:2001nq,Slavnov:2004rz,lattice-D2GC,
Xu:2011ud,Boucaud:2001st,Dudal:2003by,Dudal:2003vv,RuizArriola:2006gq,
Chernodub:2008kf,Vercauteren:2010rk}.

On the other hand, the effective Lagrangian for pure gluon system can also be constructed
in terms of lightest glueball \cite{Cornwall:1979hz,Cornwall:1982zn,Cornwall:1983zb} or one
scalar particle - dilaton \cite{Migdal:1982jp,Rosenzweig:1979ay,Dick:1998hv,Kharzeev:2002rp,Kharzeev:2008br,Chabab:2007vd}.
The dilaton field is an hypothetical scalar particle predicted by string theory and
Kaluza-Klein type theories, and its expectation value probes the strength of the gauge
coupling. In Ref.\cite{Dick:1998hv}, an effective coupling of a massive dilaton to the
4-dimensional gauge fields provides an interesting mechanism which accommodates both
the Coulomb and confining potentials between heavy quarks.

Csaki and Reece in Ref.\cite{Csaki:2006ji} proposed to model the pure gluon system
in the graviton-dilaton framework by considering the correspondence between the
dilaton background field and the non-perturbative gluon condensate, which provides a
natural IR cut-off. They investigated the case of dilaton field coupling
with dimension-4 gluon operator ${\rm Tr} \lag G^2 \rag$ and higher dimension-6
gluon condensation $ {\rm Tr} \lag G^3 \rag$, and
found that such IR correction cannot generate Regge spectra of glueball. They also
discussed a tachyon-dilaton-graviton system, where the tachyon corresponds to a
dimension-2 gluon condensate $\lag g^2A^2\rag $ to realize the linear confinement.
However, the local dimension-2 gluon operator is not gauge invariant.

In the following, we construct a 5-dimension dynamical hQCD model in the graviton-dilaton
coupled system for the pure gluodynamics, and investigate three different forms
for the dilaton background field.

\subsection{Quenched dynamical soft-wall holographic model and gluodynamics}
\label{sec-qdhm}

We construct a 5D effective model for pure gluon system by introducing one scalar dilaton
field $\Phi(z)$ in the bulk. It is not known how the dilaton field should couple with the
gauge field in 4-dimension. The 5D graviton-dilaton coupled action in the string frame is given below:
\begin{eqnarray}\label{action-graviton-dilaton}
 S_G=\frac{1}{16\pi G_5}\int
 d^5x\sqrt{g_s}e^{-2\Phi}\left(R_s+4\partial_M\Phi\partial^M\Phi-V^s_G(\Phi)\right).
\end{eqnarray}
Where $G_5$ is the 5D Newton constant, $g_s$, $\Phi$ and $V_G^s$ are the 5D
metric, the dilaton field and dilaton potential in the string frame, respectively.
The metric ansatz is often chosen to be
\begin{eqnarray}\label{metric-ansatz}
ds^2=b_s^2(z)(dz^2+\eta_{\mu\nu}dx^\mu dx^\nu), ~ ~ b_s(z)\equiv e^{A_s(z)}.
\end{eqnarray}
In this paper, the capital letters like "M,N" would stand for all the coordinates(0,1,..,4), and the greek indexes would stand for the 4D coordinates(0,...,3). We would use the convention $\eta^{00}=\eta_{00}=-1,\eta^{ij}=\eta_{ij}=\delta_{ij}$.)

Under the frame transformation
\begin{equation}
g^E_{mn}=g^s_{mn}e^{-2\Phi/3}, ~~ V^E_G=e^{4\Phi/3}V_{G}^s,
\end{equation}
Eq.(\ref{action-graviton-dilaton}) becomes
\begin{eqnarray}\label{graviton-dilaton-E}
S_G^E=\frac{1}{16\pi G_5}\int d^5x\sqrt{g_E}\left(R_E-\frac{4}{3}\partial_m\Phi\partial^m\Phi-V_G^E(\Phi)\right).
\end{eqnarray}
The equations of motion can be easily derived by doing functional variation with respective
to the corresponding fields. It takes the familiar form in the Einstein frame,
\begin{eqnarray}
E_{mn}+\frac{1}{2}g^E_{mn}\left(\frac{4}{3}\partial_l\Phi\partial^l\Phi+V_G^E(\Phi)\right)
-\frac{4}{3}\partial_m\Phi\partial_n\Phi=0,
\end{eqnarray}
and
\begin{eqnarray}
 \frac{8}{3\sqrt{g_E}}\partial_m(\sqrt{g_E}\partial^m\Phi)-\partial_{\Phi}V_G^E(\Phi)=0.
\end{eqnarray}

Under the metric ansatz Eq.(\ref{metric-ansatz}), the above Einstein equations has
two independent equations,
\begin{eqnarray}
-A_E^{''}+A_E^{'2}-\frac{4}{9}\Phi^{'2}=0, \\
\label{AEPhi}
\Phi^{''}+3A_E^{'}\Phi^{'}-\frac{3}{8}e^{2A_E}\partial_\Phi V_G^E(\Phi)=0.
\label{AEVPhi}
\end{eqnarray}
in the new variables of
\begin{equation}
b_E(z)=b_s(z)e^{-\frac{2}{3}\Phi(z)}=e^{A_E(z)},~~A_E(z)=A_s(z)-\frac{2}{3}\Phi(z).
\end{equation}
In the string frame, the above two equations of motion become
\begin{eqnarray}
-A_s^{''}-\frac{4}{3}\Phi^{'}A_s^{'}+A_s^2+\frac{2}{3}\Phi^{''}=0, \\
\Phi^{''}+(3A_s^{'}-2\Phi^{'})\Phi^{'}-\frac{3}{8}e^{2A_s-\frac{4}{3}\Phi}\partial_\Phi (e^{\frac{4}{3}\Phi}V_G^s(\Phi))=0.
\end{eqnarray}

\subsubsection{Dimension-4 dilaton background field}

As Csaki and Reece proposed proposed in \cite{Csaki:2006ji} to model the pure gluon
system in the graviton-dilaton framework by considering the correspondence between the
dilaton background field and the non-perturbative and gauge invariant dimension-4 gluon
condensate, which provides a natural IR cut-off.

Assuming the dimension-4 gluon condensate dominant in the IR region, we take the quartic
dilaton field as
\begin{equation}
\Phi(z)=\mu_{G^2}^4 z^4,
\end{equation}
and from Eq.(\ref{AEVPhi}), we can solve out the metric background and the dilaton
potential as follows:

\begin{eqnarray}
 A_{E}(z) & = &\log(\frac{L}{z})-\log(_0F_1(9/8,\frac{\Phi^2}{9})),\label{phi4AE} \\
 V_G^{E}(\Phi)& = &-\frac{4 \left(9~ _0F_1\left(\frac{1}{8},\frac{\Phi ^2}{9}\right)^2-16 \Phi ^2~ _0F_1\left(\frac{9}{8},\frac{\Phi ^2}{9}\right)^2\right)}{3 L^2}.
\end{eqnarray}

The UV expansion of the above potential is
\begin{equation}
V_G^{E}(\Phi)=-\frac{12}{L^2}+O(\Phi^4),
\end{equation}
which means the 5D mass is zero. From the AdS/CFT dictionary
$\Delta(\Delta-4)=M^2_{\Phi}L^2$, one can derive
its dimension $\Delta=4$, so it could be dual to the gauge invariant
dimension-4 gluon condensate $<g^2 {\rm G}^2>$.

As discussed in \cite{Csaki:2006ji}, and will also be shown in Sec.
\ref{glueball-quatic}, that with dimension-4 correction at IR one
cannot generate the Regge spectra for the glueball.

\subsubsection{Dimension-2 dilaton background field}

To realize the Regge behavior for the vector meson, it has been shown in
Ref. \cite{Karch:2006pv} that a quadratic dilaton background is essential.
The simplest dimension-2 dilaton background field has the quadratic form as
\begin{equation}
\Phi=\pm \mu_G^2 z^2.
\label{dilaton}
\end{equation}
The positive quadratic dilaton background is the same as the one introduced in
the KKSS model \cite{Karch:2006pv}. We will show in Section
\ref{sec-G-Glueball} and \ref{sec-G-HQ} that only positive quadratic correction
can generate the linear confinement.

In the original soft-wall model or the KKSS model \cite{Karch:2006pv}, the dilaton
field is introduced to generate the linear Regge spectra of vector meson but the
metric remains as AdS$_5$. In the graviton-dilaton coupled framework, the quadratic
dilaton field is introduced dynamically in correspondence with non-perturbative
gluodynamics, and the metric structure is automatically deformed by selfconsistently
solving the Einstein equations. With the quadratic dialton background given in
Eq.(\ref{dilaton}), we can solve the metric $A_E$ and the dilaton potential $V_G^E(\Phi)$
in the Einstein frame as
\begin{eqnarray}\label{puregluosol}
 A_E(z) & = &\log(\frac{L}{z})-\log(_0F_1(5/4,\frac{\Phi^2}{9})), \\
 V_G^E(\Phi)& = &-\frac{12 _0F_1(1/4,\frac{\Phi^2}{9})^2}{L^2}
 +\frac{16 _0F_1(5/4,\frac{\Phi^2}{9})^2\Phi^2}{3L^2},
\end{eqnarray}
with $_0F_1(a;z)$ the hypergeometric function.
It is noticed that in the Einstein frame, both the positive and negative
quadratic correction give the same results. However, in the string frame,
the positive quadratic dilaton background field $\Phi^+=\mu_G^{2}z^2$ gives:
\begin{equation}
A_s^+=A_E(z)+\frac{2}{3}\mu_G^2z^2, ~
V_G^{s,+}= e^{-4/3\mu_G^2z^2}V_{G}^E,
\label{sol-postive}
\end{equation}
and the negative dilaton background field $\Phi^-=-\mu_G^{2}z^2$ gives
\begin{equation}
A_s^-=A_E(z)-\frac{2}{3}\mu_G^2z^2, ~
V_G^{s,-}= e^{4/3 \mu_G^{2}z^2 }V_{G}^E.
\label{sol-negative}
\end{equation}
We will show in Section \ref{sec-G-Glueball} and \ref{sec-G-HQ} that positive and
negative quadratic dilaton background will induce different results on the glueball
spectra and the quark-antiquark potential.

With the normalized variable $\Phi_N$ which is defined as
\begin{equation}
\Phi \rightarrow \sqrt{\frac{3}{8}}\Phi, ~~ -\frac{4}{3}\partial_M\Phi\partial^M\Phi\rightarrow -\frac{1}{2}\partial_M\Phi\partial^M\Phi,
\end{equation}
the dilaton potential in the Einstein frame takes the form of
\begin{equation}
V_G^E(\Phi)=-12~\frac{ _0F_1(1/4;\frac{\Phi^2}{24})^2}{L^2}
+2~ \frac{ _0F_1(5/4;\frac{\Phi^2}{24})^2\Phi^2}{L^2},
\end{equation}
here $L$ the radius of AdS$_5$ and $_0F_1(a;z)$ the hypergeometric function.
In the ultraviolet limit,
\begin{equation}
V^E_G\overset{\Phi\rightarrow 0}{\longrightarrow}
-\frac{12}{L^2}+\frac{1}{2}M^2_{\Phi}\Phi^2,
\end{equation}
with the 5D mass for the dilaton field
\begin{equation}
M^2_{\Phi}L^2=-4.
\end{equation}
From the AdS/CFT dictionary $\Delta(\Delta-4)=M^2_{\Phi}L^2$, one can derive
its dimension $\Delta=2$.

It will be shown in Sec. \ref{glueball-quadratic} and \ref{meson-quadratic-quatic} that,
with positive quadratic correction to the dilaton background filed at IR, by
self-consistently solving the graviton-dilaton framework for the pure gluon system
and the graviton-dilaton-scalar framework for two-flavor system, one can produce the
scalar glueball spectra and meson spectra in good agreement with lattice/experiment
data. This indicates that some form of dimension-2 gluon operator plays important
role in QCD vaccum. Indeed, in last decade, there have been growing interests in
dimension-2 gluon condensates $\lag g^2
A^2\rag $ in SU$(N_c)$ gauge theory and its possible relation to confinement
\cite{Celenza:1986th,Lavelle:1988eg,Lavelle:1992yh,Gubarev:2000eu,Verschelde:2001ia,
Chetyrkin:1998yr,Gubarev:2000nz,Kondo:2001nq,Slavnov:2004rz,lattice-D2GC,
Xu:2011ud,Boucaud:2001st,Dudal:2003by,Dudal:2003vv,RuizArriola:2006gq,
Chernodub:2008kf,Vercauteren:2010rk}.

\subsubsection{Dilaton field with quartic form at UV and quadratic form at IR}

To build a holographic dual to the pure gluon system, we have to find the dual bulk dilaton
field which encodes the non-perturbative QCD gluodynamics. The natural candidate is the
quartic dilaton field which is dual to the gauge invariant dimension-4 gluon condensate.
Unfortunately, as shown in \cite{Csaki:2006ji} as well as in Sec. \ref{glueball-quatic},
one cannot produce confinement property of the glueball spectra with quartic dilaton field.
On the other hand, the studies in \cite{Karch:2006pv} and in Secs. \ref{glueball-quadratic}
and \ref{meson-quadratic-quatic} show that the quadratic correction to the dilaton field
at IR is essential to produce the glueball and meson spectra as well as to realize
the linear confinement. However, the gluon operator corresponding to the dimension-2
dilaton field is not well defined.

\begin{enumerate}

\item
The dimension-2 dilaton field might be dual to the dimension-2 gluon condensate $\lag g^2A^2\rag $ \cite{Gubarev:2000eu,Gubarev:2000nz,Kondo:2001nq}, which has been
discussed in some literatures, e.g. Refs.\cite{Csaki:2006ji,Andreev:2006ct,Gherghetta-Kapusta-Kelley,Afonin:2010hn}.
The simplest dimension-2 gluon operator is the zero momentum mode of $\lag g^2A^2\rag $, i.e. $<g^2A^2(k=0)>$, the Bose-Einstein condensation (BEC) of
the ``pairing" of two gluons in the vacuum due to the strong interaction \cite{Celenza:1986th,Xu:2011ud}. The BEC of the "pairing" of two gluons spontaneously generates an effective gluon mass and breaks scale invariance, and in this scenario, the dimension-4 gluon condensation is proportional to the dimension-2 gluon condensation.
Recent lattice results support a gluon mass at IR \cite{Binosi:2010ms,lattice-gluonpropagator,Dudal:2008sp} which was proposed by Cornwall
in 1981 \cite{Cornwall:1981zr} and recently developed in \cite{Kondo:2012ac}.
However, the dimension-2 gluon condensate $\lag g^2A^2\rag $ encounters the gauge
invariant problem.

\item
Motivated by Refs.\cite{Cornwall:1983zb} and \cite{Dick:1998hv},
one might introduce the holography dictionary as $\Phi^2(z)$ dual to the gauge invariant dimension-4 gluon condensation ${\rm Tr}\lag G^2 \rag $. In this case, though the dilaton field $\Phi(z)$ itself has dimension of 2, the action is always in terms of $\Phi^2$ thus there is no gauge invariant problem. However, a composite bulk operator is not consistent
gauge/gravity duality.

\item
The dimension-2 dilaton field might also correspond to the gauge invariant but non-local operator related to topological defects in the QCD vacuum \cite{Gubarev:2000nz}. However,
gauge/gravity duality requires to map a local bulk field to a local operator at the boundary.

\end{enumerate}

To avoid the gauge non-invariant problem and to meet the requirement of gauge/gravity duality,
we take the dilaton field in the form of
\begin{equation}
\Phi(z)=\mu_G^2z^2\tanh(\mu_{G^2}^4z^2/\mu_G^2).
\label{mixed-dilaton}
\end{equation}
In this way, the dilaton field at UV behaves
\begin{equation}
\Phi(z)\overset{z\rightarrow0}{\rightarrow} \mu_{G^2}^4 z^4,
\end{equation}
and is dual to the dimension-4 gauge invariant gluon condensate ${\rm Tr}\lag G^2 \rag $,
while at IR it takes the quadratic form
\begin{equation}
\Phi(z)\overset{z\rightarrow\infty}{\rightarrow} \mu_G^2 z^2,
\end{equation}
from the constraint of the linear confinement.

The dilaton potential and deformed metric can be solved numerically,
and the results on glueball spectra and meson spectra will be shown
in Secs. \ref{glueball-quadratic-quatic} and \ref{meson-quadratic-quatic}.

\subsection{Scalar glueball spectrum in quenched dynamical soft-wall model}
\label{sec-G-Glueball}

The glueball spectrum has attracted much attention more than three
decades \cite{Glueball-first}. The study of particles like
glueballs where the gauge field plays a more important dynamical
role than that in the standard hadrons, offers a good opportunity of
understanding the nonperturbative aspects of QCD, e.g. see reviews
\cite{Glueball-Review}. In Table \ref{Lat-glueballspectra}, we list
the scalar glueball spectra from several lattice groups \cite{Meyer:2004gx,Lucini:2001ej,Morningstar:1999rf,Chen:2005mg}.

The glueball has been studied in the holographic QCD models \cite{Colangelo:2007pt,Forkel:2007ru,BoschiFilho:2012xr,
Ghoroku:2011jb}.
The scalar glueball $\mr{G}$ is associated with the local gauge-invariant QCD
operator $tr(G_{\mu\nu}G^{\mu\nu})$ defined on the boundary spacetime, which has
dimension $\Delta_{\mr{G}} = 4$. From the AdS/CFT dictionary, the scalar
glueball has zero 5D mass, i.e. $M_{\mr{G},5}^2 = 0$.

We assume the glueball can be excited from the QCD vacuum described by the
quenched dynamical holographic model in Section \ref{sec-qdhm}, and
the 5D action for the scalar glueball $\mr{G}(x,z)$ in the string frame
takes the form as that in the original soft-wall model \cite{Colangelo:2007pt,Forkel:2007ru}
\begin{eqnarray}
S_{\mr{G}}=\int d^5 x \sqrt{g_s}\frac{1}{2}e^{-\Phi}\big[ \partial_M \mr{G}\partial^M
\mr{G}+M_{\mr{G},5}^2 \mr{G}^2\big].
\end{eqnarray}
The only difference is that the metric structure in the original soft-wall model is
${\rm AdS}_5$, but in our dynamical soft-wall model the metric structure is selfconsistently
solved from Section \ref{sec-qdhm}.

\begin{table}
\begin{center}
\begin{tabular}{cccccccc}
\hline\hline
n($0^{++}$) & ~Lat1   &   Lat2 &   Lat3  &   Lat4 &    Lat5            \\
\hline
  ~&  $N_c=3$  &$N_c=3$  & $N_c\rightarrow \infty $  &$N_c=3$  &$N_c=3$  \\
   1 & $1475(30)(65)$     & 1580(11)   &   1480(07)   &1730(50)(80)  &1710(50)(80)\\
   2 & $2755(70)(120)$   & 2750(35)   &   2830(22)  &2670(180)(130)   &   \\
   3 & $3370(100)(150)$      & ~        &        &       &      \\
   4 & $3990(210)(180)$      & ~        &        &       &      \\
\hline
\end{tabular}
\caption{Lattice data for $0^{++} glueball$ in unit of ${\rm MeV}$. Lat1 data
from Ref.\cite{Meyer:2004gx}, Lat2 and Lat3 data from Ref.\cite{Lucini:2001ej},
Lat4 \cite{Morningstar:1999rf} and Lat5 \cite{Chen:2005mg} are anisotropic results.}
\label{Lat-glueballspectra}
\end{center}
\end{table}

The Equation of motion for $\mr{G}$ has the form of
\begin{eqnarray}
-e^{-(3A_s-\Phi)}\partial_z(e^{3A_s-\Phi}\partial_z\mr{G}_n)=m_{\mr{G},n}^2 \mr{G}_n.
\end{eqnarray}
After the transformation $\mr{G}_n \rightarrow e^{-\frac{1}{2}(3A_s-\Phi)}\mr{G}_n$,
we get the schrodinger like equation of motion for the scalar glueball
\begin{eqnarray}
-\mr{G}_n^{''}+V_{\mr{G}} \mr{G}_n=
m_{\mr{G},n}^2 \mr{G}_n,
\label{EOM-glueball}
\end{eqnarray}
with the 5D effective schrodinger potential
\begin{equation}
V_{\mr{G}}=\frac{3A_s^{''}-\Phi^{''}}{2}+\frac{(3A_s^{'}-\Phi^{'})^2}{4}.
\label{potential-glueball}
\end{equation}

\subsubsection{Glueball spectra in original soft-wall model }
\label{sec-gb-sw}

In the original soft-wall model for glueball \cite{Colangelo:2007pt,Forkel:2007ru},
the dilaton background takes the quadratic form $\Phi=\mu_G^2z^2$ but the metric
structure is still ${\rm AdS}_5$, one can easily derive the Regge spectra for
scalar glueball:
\begin{equation}
m_{\mr{G},n}^{SW,2}= 4 \mu_G^2 (n+1), ~ n=1,2, \cdots
\label{glueballmass-AdS}
\end{equation}
which implying that the Regge slope for the scalar glueball is $4 \mu_G^2$, and the
lightest glueball mass square is $8 \mu_G^2$. In Table \ref{glueballspectra-softwall},
we list some numerical results for the scalar glueball based on
Eq. (\ref{glueballmass-AdS}) with $\mu_G=0.43, 0.6, 1 {\rm GeV}$, respectively.
\begin{table}
\begin{center}
\begin{tabular}{cccccccc}
\hline\hline
n($0^{++}$) & Soft-wall model \\
\hline
~ &$\mu_G=430$     &$\mu_G=600$      &$\mu_G=1000$            \\

        1  & 1216  & 1697   &   2828       \\
        2  & 1490  & 2078   &   3464        \\
        3   & 1720  & 2400  &   4000        \\
        4  & 1923  & 2683   &   4472      \\
\hline
\end{tabular}
\caption{ $0^{++}$ glueball in the original soft-wall model.
The unit is in ${\rm MeV}$.}
\label{glueballspectra-softwall}
\end{center}
\end{table}

From the lattice data for the scalar glueball as given in Table \ref{Lat-glueballspectra}, one can read that the slope of the Regge spectra is around $4 {\rm GeV}^2$, which means $\mu_G\simeq 1 {\rm GeV}$. From Eq. (\ref{glueballmass-AdS}), the lightest scalar glueball mass square in the soft-wall model should be around $m_{\mr{G},n=1}^{2,AdS_5} \simeq 8 {\rm GeV}^2$, which
is too large comparing with the lattice result $m_{\mr{G},n=1}^{2} \simeq 2 \sim 3 {\rm GeV}^2$.
If one fixes the lightest scalar glueball mass square $m_{\mr{G},n=1}^{2,Lat} \simeq 2 \sim 3 {\rm GeV}^2$, which gives $\mu_G \simeq 0.5 ~{\rm GeV}$, then
the slope for the Regge spectra will be around $1~ {\rm GeV}^2$, which is too small
comparing with the lattice results $4 ~{\rm GeV}^2$.

In summary, by using the ${\rm AdS}_5$ metric, the soft-wall model with the quadratic
dilaton background field cannot accommodate both the lightest scalar glueball mass and
the Regge slope.

\subsubsection{Glueball spectra with quartic dilaton background}
\label{glueball-quatic}

In the previous subsection we have shown that the positive quadratic dilaton background can generate the linear Regge behavior of $0^{++}$ glueball spectra, which agrees well with
the Lattice data \cite{Meyer:2004gx,Lucini:2001ej,Morningstar:1999rf,Chen:2005mg}. However,
dimension-4 gluon condensate is one of the most important gauge invariant non-perturbative
quantity in the QCD vacuum, it is worthwhile to investigate how much the dimension-4 gluon condensate contribute to the linear Regge behavior of the glueball spectra. Actually, in the dynamical hard-wall model\cite{Csaki:2006ji}, Csaki and Reece have studied the effect of dimension-4 gluon condensate
dual to a quartic dilaton field to mimic the IR brane effect, and they have found
that the $0^{++}$ glueball spectrum is non-linear with $m_n^2 \backsim n^2$.

\begin{table}
\begin{center}
\begin{tabular}{cccccccc}
\hline\hline
        n($0^{++}$) &  $\Phi=\mu_{G^2}^{4}z^4$ \\
\hline
~ &$\mu_{G^2}=650$  &$\mu_{G^2}=800$      \\

        0  & 1450   & 1784    &                \\

        1  & 3083   & 3795    &                 \\

        2   & 4297  & 5289     &                 \\

        3  & 5388   & 6632     &              \\
\hline
\end{tabular}
\caption{$0^{++}$ glueball spectra in the dynamic soft-wall model with quartic dilaton background $\Phi=\mu_{G^2}^{4}z^4$ in unit of ${\rm MeV}$.}
\label{glueballspectra-dimension4}
\end{center}
\end{table}

\begin{figure}[!htb]
\begin{center}
\includegraphics[width=0.5\columnwidth]{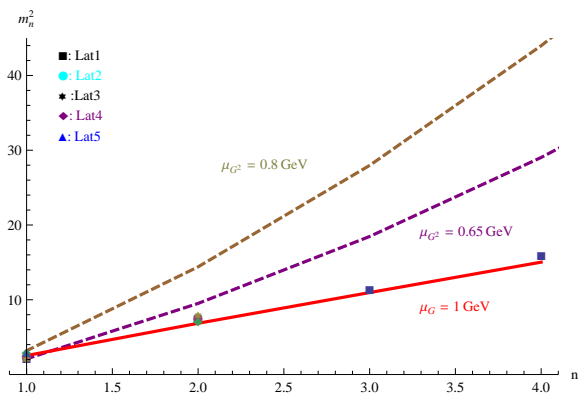}
\caption{$0^{++}$ glueball in the dynamical soft-wall with quartic background
$\Phi=\mu_{G^2}^4 z^4$ with $\mu_{G^2}=0.65, 0.8 ~{\rm GeV}$. The unit is in ${\rm GeV}$,
the dots are lattice data and the solid line is the result for quadratic background
$\Phi=\mu_{G}^2 z^2$ with $\mu_{G}=1~{\rm GeV}$. }
\label{d4glueball}
\end{center}
\end{figure}

In this subsection, we follow the approach introduced in previous subsections and study
the effect of dimension-4 gluon condensate on the glueball spectra. Then with the metric warp factor $A_{s}=A_{E}+\frac{2}{3}\Phi$ as given in Eq.(\ref{phi4AE}), we can get the effective potential in Eq.(\ref{potential-glueball}) for the glueball in this background. By solving the schrodinger-like equation with this potential, we can get the scalar glueball spectra as shown in Table \ref{glueballspectra-dimension4} and in Fig.\ref{d4glueball}.
We have chosen two sets of parameters $\mu_{G^2}=0.65 {\rm GeV}$ and $\mu_{G^2}=0.8 {\rm GeV}$ corresponding to a ground state scalar glueball mass of $m_G=1.45 {\rm GeV}$ and $m_G=1.784 {\rm GeV}$ , which are around the lightest and heaviest $0^{++}$ glueball ground state mass in Table.\ref{Lat-glueballspectra}, respectively. It is shown in Fig.\ref{d4glueball} that for both cases,
higher excitation states deviate from the linear behavior. Our result is consistent with
the result in \cite{Csaki:2006ji}, i.e. the spectra are non-linear and behave as $m_n^2 \backsim n^2$ for high excitation states. Both Ref.\cite{Csaki:2006ji} and our results show that the
quartic dilaton field which dual to the dimension-4 gluon condensate would induce the
nonlinear excitation spectra for scalar glueball.


\subsubsection{Glueball spectra with quadratic dilaton background}
\label{glueball-quadratic}

For the quadratic dilaton background field, we firstly investigate the scalar glueball
spectra with the positive quadratic dilaton background Eq.(\ref{sol-postive}).

Under the boundary condition $\mr{G}_n(0)\rightarrow 0$ and $\mr{G}_n^{'}(\infty)
\rightarrow 0$, we get the scalar glueball spectra as shown in
Table \ref{glueballspectra}. It is observed that with
$0.9 ~{\rm GeV} <\mu_G <1.1~ {\rm GeV}$, the scalar glueball spectra in the
dynamical soft-wall model with positive quadratic dilaton background
can fit lattice results quite well.
\begin{table}
\begin{center}
\begin{tabular}{cccccccc}
\hline\hline
        n($0^{++}$) &  $\Phi=\mu_{G}^{2}z^2$ \\
\hline
~ &$\mu_{G}=900$  &$\mu_{G}=1000$   &$\mu_{G}=1100$     \\

        0  & 1434   & 1593    &   1752              \\

        1  & 2356   & 2618     &   2880              \\

        2   & 2980  & 3311     &   3642              \\

        3  & 3489   & 3877     &   4264              \\
\hline
\end{tabular}
\caption{$0^{++}$ glueball spectra in the soft-wall model with positive
quadratic dilaton background $\Phi=\mu_{G}^{2}z^2$ in unit of ${\rm MeV}$.}
\label{glueballspectra}
\end{center}
\end{table}

\begin{figure}[!htb]
\begin{center}
\includegraphics[width=0.5\columnwidth]{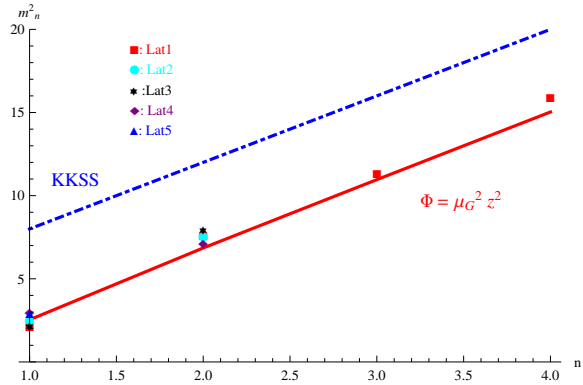}
\caption{The $0^{++}$ glueball spectra for $\Phi=\mu_G^2z^2$ with
$\mu_G=1 {\rm GeV}$ in the soft-wall model (blue dash-dotted line)
and the dynamical soft-wall model (red solid line) and compare with lattice data.
\label{glueballspectra-dilaton}}
\end{center}
\end{figure}

We would like to emphasize that the dynamical soft-wall
model has the same parameters as the original soft-wall model, i.e. the ${\rm AdS}_5$
radius $L$ which is taken to be $1$, and the quadratic coefficient of the dilaton
background field $\mu_G$. As we have shown in Sec.\ref{sec-gb-sw}, the original soft-wall
model cannot accommodate both the ground state and the Regge slope. However, if one self-consistently solves the metric background under the dynamical dilaton field, it
gives the correct ground state and at the same time gives the correct Regge slope.
This is a surprise result! To explicitly see the difference, we show the scalar glueball
spectra in the soft-wall model (blue dash-dotted line) and the dynamical soft-wall model
(red solid line) in Fig. \ref{glueballspectra-dilaton} for the case of $\mu_G=1 {\rm GeV}$.

It is observed from Fig. \ref{glueballspectra-dilaton} that the glueball spectra in the
dynamical soft-wall model is parallel to that in the soft-wall model, and the separation
is about $5.8 \mu_G^2 $. This indicates the ground sate of the scalar glueball has
mass square around $ m_{\mr{G},n=1}^2=2.5 \mu_G^2 $, and has mass around
$ m_{\mr{G},n=1}=\sqrt{2.5} \mu_G $. From numerical results, we extract the Regge
spectra in the dynamical soft-wall (DSW) model:
\begin{equation}
m_{\mr{G},n}^{2,DSW}=4 \mu_G^2 (n-1) + 2.5 \mu_G^2, ~ n=1,2, \cdots
\end{equation}

\begin{figure}[!htb]
\begin{center}
\includegraphics[width=0.48\columnwidth]{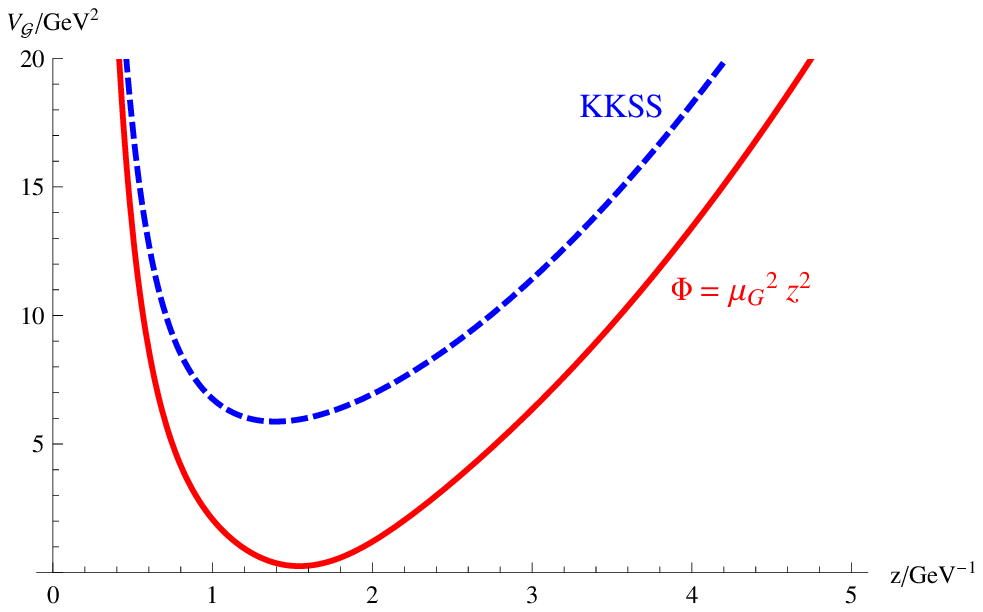}
\includegraphics[width=0.48\columnwidth]{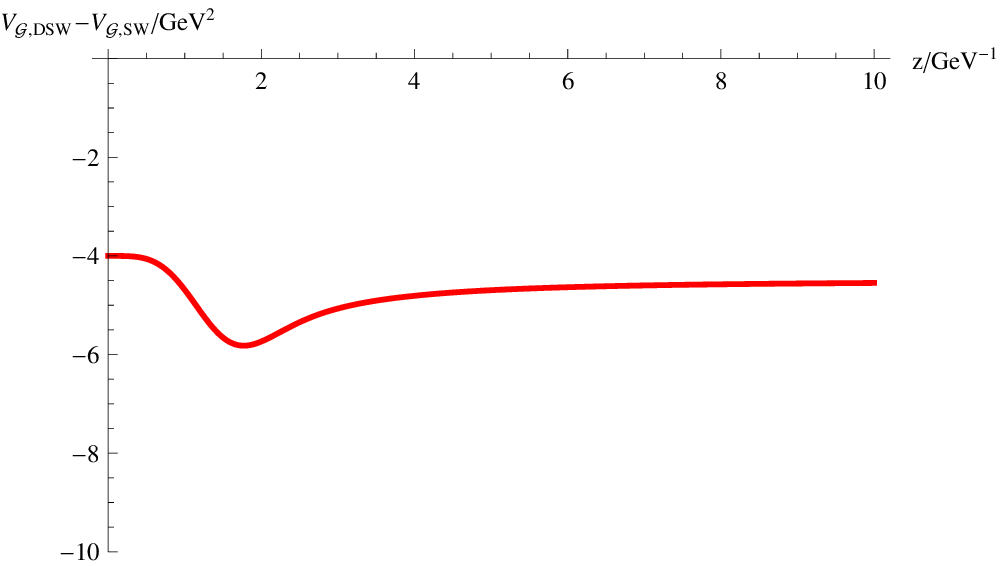}
\caption{The potentials and their difference.
\label{potentials}}
\end{center}
\end{figure}

In order to understand the difference between the soft-wall model and the dynamical
soft-wall model, we plot the effective schrodinger potentials $V_\mr{G}$
of the two models and their difference in Fig. \ref{potentials}. It is observed that
the schrodinger potential $V_\mr{G}$ (red solid line) in the dynamical soft-wall model
has a lower minimum than that in the soft-wall model (blue dashed line), the difference
is about $5.8 \mu_G^2$, which is the same as the difference of the mass square in these
two models, i.e.
$V_{\mr{G},DSW}-V_{\mr{G},SW}=m_{\mr{G},SW}^{2}-m_{\mr{G},DSW}^{2}=5.8 \mu_G^2$.

\begin{figure}[!htb]
\begin{center}
\includegraphics[width=0.5\columnwidth]{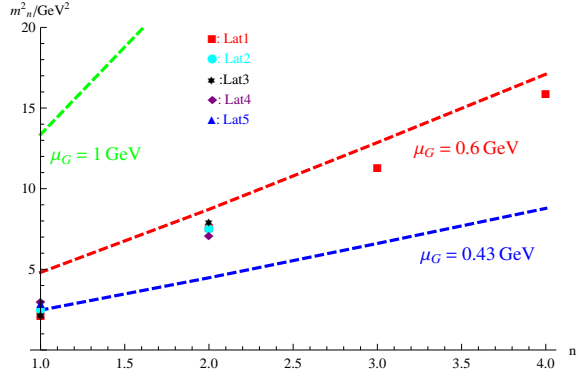}
\caption{$0^{++}$ glueball in the dynamical soft-wall model with negative quadratic background
$\Phi=-\mu_G^2 z^2$ with $\mu_G=0.43, 0.6, 1 {\rm GeV}$. The unit is in ${\rm GeV}$ and
the dots are lattice data.}
\label{glueball-negative}
\end{center}
\end{figure}

If the dynamical soft-wall model takes the negative quadratic dilaton background
$\Phi=-\mu_G^2 z^2$, the metric structure has the form of Eq.(\ref{sol-negative}),
and the scalar glueball spectra is shown in Fig. \ref{glueball-negative} with
$\mu_G=0.43, 0.6, 1 {\rm GeV}$, respectively.

It is observed that the negative quadratic dilaton background can also generate
the Regge spectra. However, like the soft-wall model, the dynamical
soft-wall model with negative dilaton cannot accommodate both the ground
state and the Regge slope.

\subsubsection{Glueball spectra for dilaton field with quartic form at UV
and quardratic form at IR}
\label{glueball-quadratic-quatic}

\begin{figure}[!htb]
\begin{center}
\includegraphics[width=0.5\columnwidth]{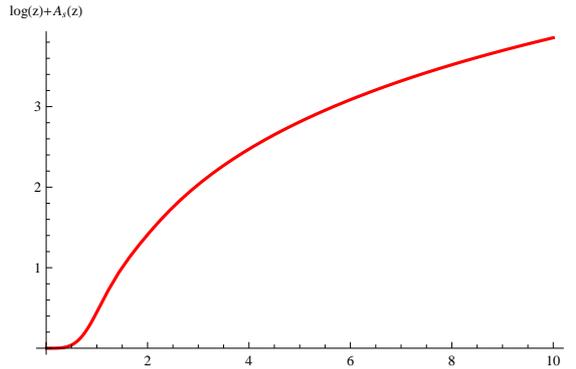}
\caption{The deformed metric $A_s$ as function of $z$ for the dilaton field
$\Phi(z)=\mu_G^2z^2\tanh(\mu_{G^2}^4z^2/\mu_G^2)$
with $\mu_G=\mu_{G^2}=1 {\rm GeV}$. Here we plot $\log(z)+A_s(z)$
to avoid the $\log(z)$ divergence at $z=0$ of $A_s$ which comes from the
approximate AdS behavior of the solution.}
\label{z4-z2As}
\end{center}
\end{figure}

For the dilaton background field Eq.(\ref{mixed-dilaton}) with quartic form at UV
and quardratic form at IR, we can solve the background metric under this dilaton field
from the equation of motion Eq.(\ref{AEVPhi}), and the numerical result is shown in Fig.\ref{z4-z2As}.

\begin{figure}[!htb]
\begin{center}
\includegraphics[width=0.5\columnwidth]{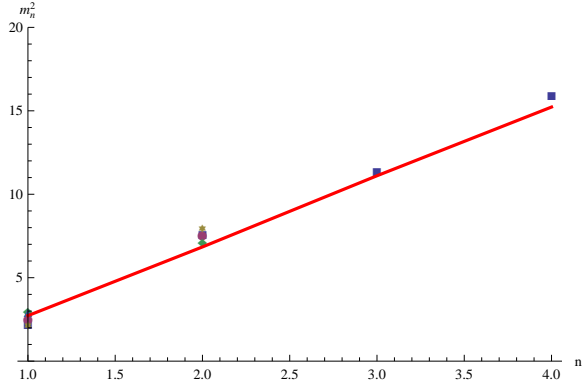}
\caption{Glueball spectra for the dilaton background
$\Phi(z)=\mu_G^2z^2\tanh(\mu_{G^2}^4z^2/\mu_G^2)$
with $\mu_G=\mu_{G^2}=1 {\rm GeV}$.}
\label{z4-z2glueball}
\end{center}
\end{figure}

Then from Eq. (\ref{EOM-glueball}), we can solve the scalar glueball spectra as in the
previous sections and the result is shown in Fig.\ref{z4-z2glueball}. It is found that
the glueball spectra is not sensitive to the value of $\mu_{G^2}$ as long as
$\mu_{G^2}>\mu_G$. For $\mu_G=\mu_{G^2}=1 {\rm GeV}$,
the scalar glueball spectra for the dilaton field
$\Phi(z)=\mu_G^2z^2\tanh(\mu_{G^2}^4z^2/\mu_G^2)$ is almost the same
as that for the quadratic dilaton field
$\Phi(z)=\mu_G^2 z^2$ with $\mu_G=1 {\rm GeV}$.

\subsection{Linear quark potential in quenched dynamical soft-wall model}
\label{sec-G-HQ}

We follow the standard procedure \cite{Maldacena:1998im,Rey:1998bq} to
derive the static heavy quark potential $V_{Q{\bar Q}}(r)$ in the dynamical
soft-wall holographic model under the general metric background Eq.(\ref{metric-ansatz}). In ${\rm SU(N)}$ gauge theory, the interaction
potential for infinity massive heavy quark antiquark is calculated
from the Wilson loop
\begin{equation}
W[C]=\frac{1}{N} {\rm Tr} P \exp[i \oint_{C} A_\mu dx^\mu],
\label{Wilson-loop-formula}
\end{equation}
where $A_{\mu}$ is the gauge field, the trace is over the
fundamental representation, $P$ stands for path ordering. $C$
denotes a closed loop in space-time, which is a rectangle with one
direction along the time direction of length $T$ and the other space
direction of length $R_{Q\bar Q}$.

The Wilson loop describes the creation of a
$Q{\bar Q}$ pair with distance $R$ at time $t_0=0$ and the
annihilation of this pair at time $t=T$. For $T\to\infty$, the
expectation value of the Wilson loop behaves as $\langle
W(C)\rangle\propto e^{-T V_{Q\bar Q}}$. According to the
\textit{holographic} dictionary, the expectation value of the Wilson
loop in four dimensions should be equal to the string partition
function on the modified ${\rm AdS}_5$ space, with the string world
sheet ending on the contour $C$ at the boundary of ${\rm AdS}_5$
\begin{equation}
\langle W^{4d}[C]\rangle=Z_{string}^{5d}[C]\simeq e^{-S_{NG}[C]} \,\ ,
\end{equation}
where $S_{NG}$ is the classical world sheet Nambu-Goto action
\begin{equation} \label{S-NG-HQ}
S_{NG}=\frac{1}{2\pi\alpha_p}\int d^2 \eta \sqrt{{\rm Det} \chi_{a
b}},
\end{equation}
with $\alpha_p$ the 5D string tension which has dimension of ${\rm
GeV}^{-2}$, and $\chi_{ab}$ is the induced worldsheet metric with
$a,b$ the two indices of the world sheet coordinates ($\eta^0,\eta^1$).
Without loss of generality, we can choose the $\eta^0=t,\eta^1=x$,
and the position of one quark is $x=-\frac{R_{Q\bar Q}}{2}$ and the other is
$x=\frac{R_{Q\bar Q}}{2}$. Under the background (\ref{metric-ansatz}),
the Nambu-Goto action Eq.(\ref{S-NG-HQ}) becomes
\begin{equation} \label{S-NG-BG}
S_{NG}=\frac{T L^2}{2\pi\alpha_p}\int dx  e^{2A_s}\sqrt{1+z^{'2}},
\end{equation}
with the prime $'$ denotes the derivative with respective to $x$.

Since there's no dependence on $x$, we can easily obtain the equation of
motion:
\begin{equation}\label{equa-HQP-g}
\frac{e^{2\mathcal
{A}_s(z)}}{\sqrt{1+(z')^2}}= \text{Constant}=
e^{2\mathcal{A}_s(z_0)},
\end{equation}
for the minimum world-sheet surface configuration.

Here the $R_{Q\bar Q}$ is dependent on $z_0$ which is the maximal value of $z$
and $ z'(x=0)=0$. For the configuration mentioned above and the given
equation of motion, we impose the following boundary condtions $
z(x=0)=z_0,
 z(x=\pm \frac{R_{qq}}{2})=0$. Following the standard procedure, one can
derive the interquark distance $R_{Q\bar Q}$ as a function of $z_0$
\begin{eqnarray}
R_{Q\bar Q}(z_0)=2\int_{0}^{z_0}
dz\frac{1}{\sqrt{1-\frac{b_s^4(z_0)}{b_s^4(z)}}}\frac{b_s^2(z_0)}{b_s^2(z)}.
\label{Rqq-g}
\end{eqnarray}
The heavy quark potential can be worked out from the Nambu-Goto
string action:
\begin{eqnarray}
V_{Q\bar Q}(z_0)=\frac{g_p}{\pi}\int_{0}^{z_0}
dz\frac{b_s^2(z)}{\sqrt{1-\frac{b_s^4(z_0)}{b_s^4(z)}}},
\label{vqq-general-g}
\end{eqnarray}
with $g_p=\frac{L^2}{\alpha_p}$. It is noticed that the integral in
Eq.(\ref{vqq-general-g}) in principle include a pole in the UV region
($z\rightarrow 0$), which
induces $ V_{Q\bar Q}(z)\rightarrow \infty$. The infinite energy
should be extracted through certain regularization procedure. The
divergence of $V_{Q\bar Q}(z)$ is related to the vacuum energy for
two static quarks. Generally speaking, the vacuum energy of two
static quarks will be different in various background. In our latter
calculations, we will use the regularized $V_{Q\bar Q}^{ren.}$ ,
which means the vacuum energy has been subtracted.
A minimal subtracted result related to the background solution
Eq.(\ref{puregluosol}) is as following,
\begin{eqnarray}
& &V_{Q\bar Q}(z_0)=\frac{g_p}{\pi z_0}(\int_{0}^{1}
d\nu(\frac{b_s^2(z_0 \nu)z_0^2}{\sqrt{1-\frac{b_s^4(z_0)}{b_s^4(z_0
\nu)}}}-\frac{1}{\nu^2})-1), ~~~ \label{vqqrn-g} \\
& & R_{Q\bar Q}(z_0)=2 z_0 \int_{0}^{1}
d\nu\frac{1}{\sqrt{1-\frac{b_s^4(z_0)}{b_s^4(z_0\nu)}}}\frac{b_s^2(z_0)}{b_s^2(z_0\nu)}.
\label{Rqq-g}
\end{eqnarray}

The integrate kernel in Eq.(\ref{vqqrn-g}) has a
pole at $\nu=1$, and by expanding the integral kernel at $\nu=1$ one has
\begin{eqnarray}\label{nu-1-g}
1-\frac{b_s^4(z_0)}{b_s^4(z_0\nu)} =
\frac{4z_0b_s^{'}(z_0)}{b_s(z_0)}(\nu-1)+o((\nu-1)^2)\,.
\end{eqnarray}
From Eqs.(\ref{vqqrn-g},\ref{Rqq-g},\ref{nu-1-g}), we can find the
necessary condition for the linear quark potential is that:
There exists a point $z_c$, at which
\begin{equation}
b_s^{'}(z_c)\rightarrow 0, b_s(z_c)\rightarrow const,
\label{criteria}
\end{equation}
then the integral is dominated by $\nu=1$ region, one can obtain the string tension
\begin{eqnarray}
\sigma_s \propto \frac{V_{Q\bar Q}(z_0)}{R_{\bar{q}q}(z_0)}\overset{z_0\rightarrow z_c}{\longrightarrow} \frac{g_p}{2\pi} b_s^2(z_c). \label{stringtension-g}
\end{eqnarray}

\begin{figure}[!htb]
\begin{center}
\includegraphics[width=0.5\columnwidth]{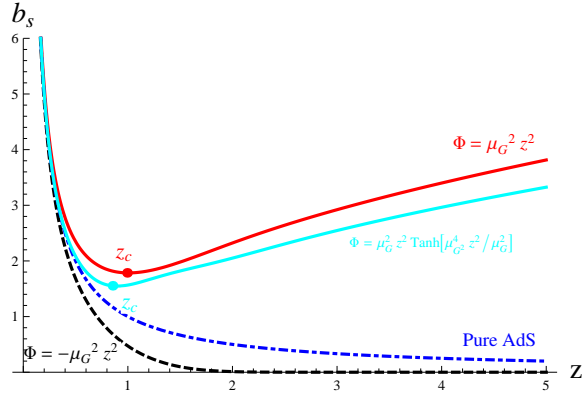}
\caption{The metric structure $b_s(z)=e^{A_s(z)}$ as functions of $z$
corresponding to $\Phi=\mu_G^2z^2$ (red solid line), $\Phi=-\mu_G^2z^2$ (black dashed line),
and $\Phi=\mu_G^2z^2\tanh(\mu_{G^2}^4z^2/\mu_G^2)$ (cyan solid line), respectively.
The blue dash-dotted line stands for the pure ${\rm AdS}_5$ case. $\mu_G=1 {\rm GeV}$ has been taken for numerical calculation.}
\label{quenched-bs}
\end{center}
\end{figure}
\begin{figure}[!htb]
\begin{center}
\includegraphics[width=0.6\columnwidth]{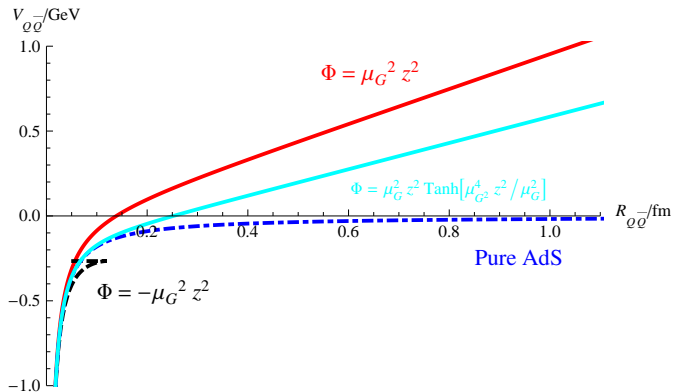}
\caption{The quenched quark potential result $V_{Q\bar Q} $ as functions of $R_{Q\bar Q}$
in the quenched dynamical soft-wall model for the dilaton field $\Phi=\mu_G^2z^2$ (red solid line), $\Phi=-\mu_G^2z^2$ (black dashed line), $\Phi=\mu_G^2z^2\tanh(\mu_{G^2}^4z^2/\mu_G^2)$ (cyan solid line), respectively.  The blue dash-dotted line stands for the pure ${\rm AdS}_5$ case.
$\mu_G=1 {\rm GeV}$ and $g_p=0.4$ have been used for numerical calculations. }
\label{vqq-quenched}
\end{center}
\end{figure}

Fig.\ref{quenched-bs} shows the metric structure $b_s(z)$ as functions of $z$ for the
${\rm AdS}_5$ metric (blue dash-dotted line), and for the solutions of the
quenched dynamical soft-wall model with dilaton background fields $\Phi=\mu_G^2z^2$
(red solid line), $\Phi=-\mu_G^2z^2$ (black dashed line) and $\Phi=\mu_G^2z^2\tanh(\mu_{G^2}^4z^2/\mu_G^2)$ (cyan solid line), respectively.
We can see that only for
the case of positive dilaton background $\Phi=\mu_G^2z^2$ and $\Phi=\mu_G^2z^2\tanh(\mu_{G^2}^4z^2/\mu_G^2)$, the metric solution
Eq.(\ref{sol-postive}) has a minimum point $z_c$. Therefore, the quark-antiquark potential
should have a linear part for positive quadratic dilaton background $\Phi=\mu_G^2z^2$ and
for $\Phi=\mu_G^2z^2\tanh(\mu_{G^2}^4z^2/\mu_G^2)$, which
can be seen explicitly from Fig.\ref{vqq-quenched}. While for the pure ${\rm AdS}_5$ case
as well as for the dynamical soft-wall model with negative dilaton background field
$\Phi=-\mu_G^2z^2$, there doesn't exist a $z_c$ where $b_s^{'}(z_c)\rightarrow 0$, and
correspondingly the heavy quark potential does not show a linear behavior at large $z$.

\subsection{Short summary}

In this section, we have modeled the pure gluon system by using the quenched dynamical
soft-wall model in the graviton-dilaton framework.
Comparing with the original soft-wall model with ${\rm AdS}_5$ metric, here the metric
background at IR is self-consistently deformed by the gluon condensate. The quartic dilaton field effect should be negligible in the confinement issue.

It is found that the positive quadratic dilaton background can give the correct glueball spectra including the Regge slope and ground state, as well as the linear quark potential, and the
negative quadratic dilaton background field can be safely excluded. In the following study, we
will only focus on the case of IR positive quadratic dilaton background.

\section{Two flavor system: KKSS model and improved KKSS model}
\label{sec-hadronspectra-KKSS}

We now turn to the the light flavor system with chiral symmetry $SU(2)_L \times SU(2)_R$.
As we have mentioned in the Introduction, the current achievements of AdS/QCD models for
hadron spectra are the hard-wall AdS/QCD model \cite{EKSS2005} and the soft-wall AdS/QCD
or KKSS model \cite{Karch:2006pv} and its extended version
\cite{Colangelo:2008us, Gherghetta-Kapusta-Kelley,YLWu,Afonin:2012jn}.
In the hard-wall model \cite{Karch:2006pv}, the chiral symmetry breaking can be realized
by chiral condensation in the vacuum, however, the resulting mass spectra for the excited
mesons behave as $m_n^2\sim n^2$, which is different from the linear Regge behavior
$m_n^2\sim n$. In order to generate the linear Regge behavior, the authors of
Ref.\cite{Karch:2006pv} introduced a quadratic dilaton background, one can obtain a
desired mass spectra for the excited vector mesons, while the chiral symmetry breaking
phenomenon cannot consistently be realized. In the following, we firstly give a brief
introduction on the KKSS model and review the meson spectra in this model.

\subsection{The KKSS model}

The KKSS model \cite{Karch:2006pv} has two background fields:
the positive quadratic dilaton background $\Phi=\mu^2z^2$ and the metric background $g_{MN}$.
Note, in the following, we will use $\mu$ instead of $\mu_G$ to distinguish from the
pure gluon system. The background geometry is not dynamically generated but assumed to be
${\rm AdS}_5$ space with the metric structure
\begin{equation}\label{equstringmetric}
ds^{2}=g_{MN}dx^{M}dx^{N}=\frac{L^2}{z^2}\left( \eta_{\mu\nu}dx^{\mu}dx^{\nu}+dz^{2}\right),
\end{equation}
which gives $A_s(z)=-\log(z/L)$.

The mesons are described by 5D fields propagating on the background with the action given by
\begin{eqnarray}
 S_{\rm KKSS}=-\int d^5x e^{-\Phi(z)} \sqrt{g_s}Tr\Big(|DX|^2+m_X^2 X^2
 +\frac{1}{4g_5^2}(F_L^2+F_R^2)\Big), \label{action-KKSS}
\end{eqnarray}
with $g_5=12\pi^2/N_c$. The scalar field $X$ is dual to the dimension-3 $q\bar{q}$ operator,
and $m_X$ is the 5D scalar mass. According to AdS/CFT dictionary, the dimension-3 scalar
has 5D mass $m_{X}^{2}=-3$.
The field $X(z)$ is actually a complex field to incorporate the scalar $S$ and
the pseudoscalar $P$ fields,
\begin{equation}
X^{\alpha\beta}(z) = \left(\frac{\chi(z)}{2}+S\right)\emph{1}^{\alpha\beta}{\rm e}^{iP^{a}t^{a}},
\label{scalarfield}
\end{equation}
where $\alpha,\beta$ are in the isospin space, $a=1,2,3$ are the $SU(2)$ generator index.
The scalar field takes a nonzero vacuum expectation
value (VEV) $\chi(z)$, which is expected to realize the chiral symmetry breaking.

The Gauge fields $L_{M}$ and
$R_{M}$ model the SU(2)$_L\times$ SU(2)$_R$ global chiral symmetry of QCD
for two flavors of quarks, which are defined as
\begin{eqnarray}
F_{L}^{MN}&=&\partial^{M}{L^{N}}-\partial^{N}{L^{M}}-i[L^{M},L^{N}],\nonumber\\
F_{R}^{MN}&=&\partial^{M}{R^{N}}-\partial^{N}{R^{M}}-i[R^{M},R^{N}],
\end{eqnarray}
where $L^{M}= L^{Ma} t^a$ and Tr$[t^{a}t^{b}]=\delta^{ab}/2$. The covariant derivative becomes
\begin{equation}
D^M X=\partial^M X-i L^M X+iX R^M.
\end{equation}

To describe the vector and axial-vector fields, we simply transform the $L$ and $R$ gauge
fields into the vector ($V$) and axial-vector ($A$) fields with
$L^{M}=V^{M}+A^{M}$ and $ R^{M}=V^{M}-A^{M}$, one can have
$F_{L}^{2}+F_{R}^{2} = 2 \left(F_{V}^{2}+  F_{A}^{2}\right)$, with
\begin{eqnarray}
F_{V}^{MN}&=&\partial^{M}{V^{N}}-\partial^{N}{V^{M}}-\frac{i}{\sqrt{2}}[V^{M},V^{N}],\\
F_{A}^{MN}&=&\partial^{M}{A^{N}}-\partial^{N}{A^{M}}-\frac{i}{\sqrt{2}}[A^{M},A^{N}].
\end{eqnarray}

In terms of the vector $V$ and axial-vector $A$ fields, the KKSS action Eq.(\ref{action-KKSS})
can be rewritten as
\begin{equation}
S_{KKSS}=-\int d^{5}x \sqrt{g_s}\,{\rm e}^{-\Phi(z)}{\rm Tr}\left[|D X|^{2}+ m_{X}^{2} |X|^{2}
+\frac{1}{2g_{5}^{2}}(F_{V}^{2}+F_{A}^{2})\right],
\label{action-KKSS-VA}
\end{equation}
where the covariant derivative now becomes
\begin{equation}
D^M X=\partial^M X-i[V^{M},X]-i\{A^{M},X\}.
\end{equation}

\subsection{Degeneration of chiral partners in KKSS model}

The scalar field takes a nonzero vacuum expectation value (VEV) $\chi(z)$,
which is expected to realize the chiral symmetry breaking as in the hard wall model.
We will show in the following that the chiral symmetry breaking is not realized in
the soft-wall model or KKSS model, and we will analyze the reason.

\vskip 0.5cm
\noindent {\bf Scalar vacuum expectation value}
\vskip 0.5cm

The equation of motion for the scalar vacuum expectation value (VEV) $\chi(z)$ defined in
Eq.(\ref{scalarfield}) can be deduced and takes the following form,
\begin{eqnarray}
\chi^{''}+(3 A_s^{'}-\Phi^{'})\chi^{'}- m_X^2 e^{2A_s}\chi=0.
\label{scalarVEV-KKSS}
\end{eqnarray}

In the hard wall model, $\Phi^{'}=0$, the scalar VEV has the exact solution
\begin{equation}
\chi(z)=c_1z+c_2z^3=m_qz+\sigma z^3,
\end{equation}
where we have identified $m_q=c_1$ and $\sigma=<{\bar q}q>=c_2$ ( As shown in \cite{Cherman:2008eh}, a normalization constant might appear between $m_q$ and $\sigma$ to match the QCD result. However this factor would not affect the main discussion in this section, so we just follow the settings in the original soft-wall model\cite{Karch:2006pv} here).
In the softwall model, $\Phi(z)=\mu^2 z^2$ and $\Phi^{'}=2\mu^2z$, and the general
solution of Eq.(\ref{scalarVEV-KKSS}) has the form of
\begin{eqnarray}
\chi(z)=c_2 G_{1,2}^{2,0}\left(-z^2|
\begin{array}{c}
 1 \\
 \frac{1}{2},\frac{3}{2}
\end{array}
\right)+c_1 e^{\frac{z^2}{2}} z^3 \left(I_0\left(\frac{z^2}{2}\right)
+I_1\left(\frac{z^2}{2}\right)\right).
\end{eqnarray}
with $I_n(z)$ the modified Bessel function of the first kind,$G_{pq}^{mn}\left(z\left|
\begin{array}{c}
 a_1,\ldots ,a_p \\
 b_1,\ldots ,b_q
\end{array}
\right.\right)$ the MeijerG function(For details, please refer to Mathematica 8.0).

To be more instructive, we can extract the large z behavior of $\chi$  from the equation of motion : assuming
$\chi^{''}<<\chi$,when $z>>1$, we have $-2\mu^2 z \chi^{'}+3\chi/z^2=0$, and assuming
$\chi^{'}>>\chi$,when $z>>1$, we have $\chi^{''}-2\mu^2 z \chi^{'}=0$. And then we could get the
IR behaviors of the two independent solution: $\chi_1 \rightarrow e^{\mu^2 z^2}/(\mu z)$
and $\chi_2 \rightarrow e^{-3/(4 \mu^2 z^2)}\rightarrow 1$.  The first one would make the spectra
of $a_1$ nonlinear, so in order to produce linear $a_1$ spectra,  $c_1,c_2$ in Eq.(\ref{scalarVEV-KKSS})
are not independent. Requiring $ \chi \propto \chi_2(z)$, when $z>>1$, the small $z$ expansion of
$\chi$ would be:
\begin{eqnarray}
\chi(z)&&=c(\mu z+\mu^3z^3 \left(-\frac{1}{2}
+\gamma_E +\frac{\psi \left(-\frac{1}{2}\right)}{2}+\log (\mu z)\right))+O(z^4)\\
&&=c(\mu z+\mu^3z^3(0.095+\log(\mu z)))
\end{eqnarray}
with $\gamma_E=0.577$ the Euler's constant and $\psi(z)$ the digmma function with $\psi(-\frac{1}{2})=0.036$.
We would take $c=m_q/\mu$, so $\sigma$ can be read as $\sigma=0.095 m_q \mu^2$.
The experimental data for vector, axialvector, scalar and pseudoscalar are shown in
Table \ref{exp-mass}. To fit the Regge slope of vector meson $\rho$, we have to choose
$\mu=0.43$. Then even we take $m_q=9 {\rm MeV}$, $\sigma$ is only $(54~{\rm MeV})^3$, which is
too small comparing with the experienced value $(250~{\rm MeV})^3$. This problem
was pointed in the original paper and the authors also mentioned to add quartic terms
$|X|^4$ to cure it.

\begin{table}
\begin{center}
\begin{tabular}{cccccccc}
\hline\hline
   Exp.   &  n & $\rho$~(MeV)  & $a_1$~(MeV)     & $f_0$~(MeV)    & $\pi$~(MeV)\\  \hline
      &  1 & $775\pm1$    &  $1230\pm40$    & $550^{+250}_{-150}$  & $140$\\
      &  2 & $1282\pm37$  &   $1647\pm22$   & $980 \pm 10$        & $1300\pm100$\\
      &  3 & $1465\pm25$  &   $1930^{+30}_{-70}$ & $1350 \pm 150$ & $1816\pm14$\\
      &  4 & $1720\pm20$  &    $2096\pm122$  & $1505 \pm 6$     & $2070$   \\
      &  5 & $1909\pm30$  &    $2270^{+55}_{-40}$  & $1724 \pm 7$  & $2360$\\
      &  6 & $2149\pm17$  &   ----           & $1992 \pm 16$     &----          \\
      &  7 & $2265\pm40$  &    ----  & $2103 \pm 8$    &----              \\
      &  8 & ----        &----       &  $2314 \pm 25$   &----    \\
\hline\hline
\end{tabular}
\caption{The experimental data for meson mass from PDG \cite{pdg}.
The data selection scenario used here is the same
as in Ref.\cite{Gherghetta-Kapusta-Kelley}, which shows the chiral
symmetry breaking maintains in the highly excited states of chiral partners.}
\label{exp-mass}
\end{center}
\end{table}

\vskip 0.2cm
\noindent {\bf Meson spectra}
\vskip 0.2cm

In the following, we show the meson spectra in the KKSS model.
The equations of motion of the vector, axial-vector, scalar and pseudo-scalar mesons
take the form of:
\begin{eqnarray}
-\rho_n^{''}+V_{\rho} \rho_n&=&m_n^2 \rho_n, \\
-a_n^{''}+ V_a a_n&=& m_n^2 a_n, \\
-s_n^{''}+V_s s_n&=&m_n^2 s_n, \\
 -\pi_n''+V_{\pi,\varphi} \pi_n & = & m_n^2(\pi_n-e^{A_s}\chi\varphi_n), \nonumber\\
 -\varphi_n''+V_{\varphi} \varphi_n & = & g_5^2 e^{A_s}\chi(\pi_n-e^{A_s}\chi\varphi_n).
\end{eqnarray}
with schrodinger like potentials
\begin{eqnarray}
V_{\rho}&=& \frac{A_s^{'}-\Phi^{'}}{2}+\frac{(A_s^{'}-\Phi^{'})^2}{4}, \\
V_a&=& \frac{A_s^{'}-\Phi^{'}}{2}+\frac{(A_s^{'}-\Phi^{'})^2}{4}+g_5^2 e^{2A_s} \chi^{2},\\
V_s&=& \frac{3A_s^{''}-\phi^{''}}{2}+\frac{(3A_s^{'}-\phi^{'})^2}{4}-m_X^2e^{2A_s}, \\
V_{\pi,\varphi}&=& \frac{3A_s^{''}-\Phi^{''}+2\chi^{''}/\chi-2\chi^{'2}/\chi^2}{2}
           +\frac{(3A_s^{'}-\Phi^{'}+2\chi^{'}/\chi)^2}{4}, \\
V_{\varphi}&=& \frac{A_s^{''}-\Phi^{''}}{2}+\frac{(A_s^{'}-\Phi^{'})^2}{4}.
\end{eqnarray}
Since $g_5^2 e^{2A_s} \chi^{2}\rightarrow 0$ when $z\rightarrow\infty$, we can
expect the spectra of the chiral partners, i.e. the vector and axial vector as well
as the scalar and pseudoscalar mesons would be degenerate in the large $n$ region.

\begin{table}
\begin{center}
\begin{tabular}{cccccccc}
\hline\hline
  KKSS   &   n & $\rho$~(MeV) & $a_1$~(MeV) &$f_0$~(MeV) & $\pi$~(MeV)\\ \hline
     &   1 & $860$   & 860 &1053   & 1054\\
     &   2 & $1216 $ & 1217   &1360   &1360\\
     &   3 & $1489$  &1490    &1609   &1609\\
     &   4 & $1720 $  &1720   &1824   &1824\\
     &   5 & $1923$   &1923   &2017   &2017\\
     &   6 & $2107$   &2107   &2193   &2192\\
     &  7 & $2275$   &2275   &2355    &2355\\
\hline\hline
\end{tabular}
\caption{The mass spectra for vector mesons $\rho$, axial vector mesons $a_1$,
scalar mesons $f_0$ and pseudoscalar mesons $\pi$ in the KKSS model
with $m_q=9 ~{\rm MeV},\mu=430~ {\rm MeV}$, which gives $\sigma=(54 ~{\rm MeV})^3$).}
\label{T-spectra-KKSS}
\end{center}
\end{table}

\begin{figure}[!htb]
\begin{center}
\includegraphics[width=0.5\columnwidth]{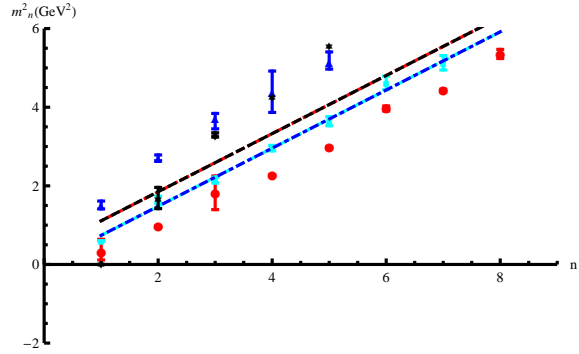}
\caption{Meson spectra in the KKSS model with $m_q=9~{\rm MeV},\mu=430~{\rm MeV}$
comparing with experimental data in Table \ref{exp-mass}.}
\label{spectra-KKSS}
\end{center}
\end{figure}

The meson spectra(solved from the equation of motion with boundary condition$\psi(0)=0,\partial_z \psi(z\rightarrow)=0$, $\psi=\rho_,a_n,s_n,\pi_n,\varphi_n$) in the KKSS model is shown in Table \ref{T-spectra-KKSS} and
Fig.\ref{spectra-KKSS}. In order to realize the linear Regge behavior,
we have used parameters $m_q=9 ~{\rm MeV},\mu=430~ {\rm MeV}$.
However, this gives a small chiral condensate $\sigma=(54 {\rm MeV})^3$, which
leads to the degeneration of chiral partners,
i.e. the scalar spectra overlaps with the pseudoscalar spectra, and the vector spectra
overlaps with the axial-vector spectra. On the other hand, in order to realize the
chiral symmetry breaking in the KKSS model, i.e. the separation
of the spectra of the chiral partners, as shown in \cite{Huang:2007fv},
one cannot get the linear Regge behavior for the axial-vector meson.

\subsection{Improved KKSS model with quartic interaction term}

As we have shown above that the KKSS model cannot accommodate chiral symmetry breaking
and linear confinement. Refs. \cite{Colangelo:2008us, Gherghetta-Kapusta-Kelley}
introduced a quartic interaction term $\kappa X^4$ in the bulk scalar potential to
improve the situation. Nevertheless, such a term was shown in
Ref. \cite{Gherghetta-Kapusta-Kelley} by Gherghetta-Kapusta-Kelley to
result in a negative mass for the lowest lying scalar meson state.

\begin{figure}[!htb]
\begin{center}
\includegraphics[width=0.5\columnwidth]{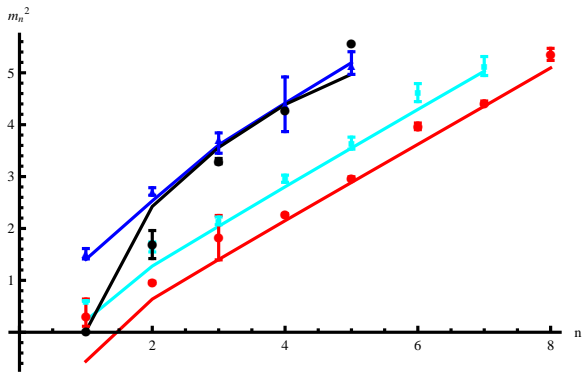}
\caption{Meson spectra in the GKK model with $m_q=9.75~{\rm MeV},\sigma=(204.5 ~{\rm MeV})^3,\mu=430~{\rm MeV}$.}
\label{all-GKK}
\end{center}
\end{figure}

The meson spectra in the Gherghetta-Kapusta-Kelley (GKK) model is shown in Fig.\ref{all-GKK},
where the parameters are chosen: $m_q=9.75 {\rm MeV},\sigma=(204.5~{\rm MeV})^3,
\mu=430~{\rm MeV}$ (equivalent to $\lambda=0.183GeV^2$ in their notation). The lowest scalar meson has mass square
$m_{f_0,n=1}^2= -0.559 {\rm GeV}^2$, which shows the instability in the scalar sector.

\subsection{Improved KKSS model with deformed warp factor}

In Ref.\cite{YLWu}, Sui-Wu-Xie-Yang (SWXY) introduced a deformed warp factor in
the KKSS and GKK model and the metric structure takes the form of
\begin{equation}
b_s(z)=\frac{1+\mu_g^2z^2}{z^2},
\label{bs-SWXY}
\end{equation}
which can cure the instability
of the scalar potential and produce meson spectra in good agreement
with experimental data. Even though the authors in Ref.\cite{YLWu}
didn't calculate the heavy quark potential, but from our criteria for
the linear quark potential Eq. (\ref{criteria}), the geometric factor
Eq.(\ref{bs-SWXY}) in the SWXY model can produce a linear potential.

In this model, the authors grouped their settings into "case-a" and "case-b".

1) In "case-a", the large z behavior of the scalar (see Eq.(9) in their paper) is
$\chi(z\rightarrow \infty)=\gamma(\mu z)$, since their metric warp factor is
like $A_s(z \rightarrow \infty) =constant$, the difference between the effective
potential in vector sector and axial-vector sector takes the limit of
$g_5^2 \chi^2 e^{2A_s}\propto z^2$, we can see the Regge slope for the vector
spectra is different from that for the axial-vector mesons.

2) In "case-b", $\chi(z\rightarrow \infty)=\gamma(\sqrt{\mu z})$, the difference
between the effective potential in vector sector and axial-vector sector takes the
limit of $g_5^2 \chi^2 e^{2A_s}\propto z$, and the vector
and axial-vector spectra would approach each other at high excitations.

\begin{figure}[h]
\begin{center}
\epsfxsize=6.5 cm \epsfysize=6.5 cm \epsfbox{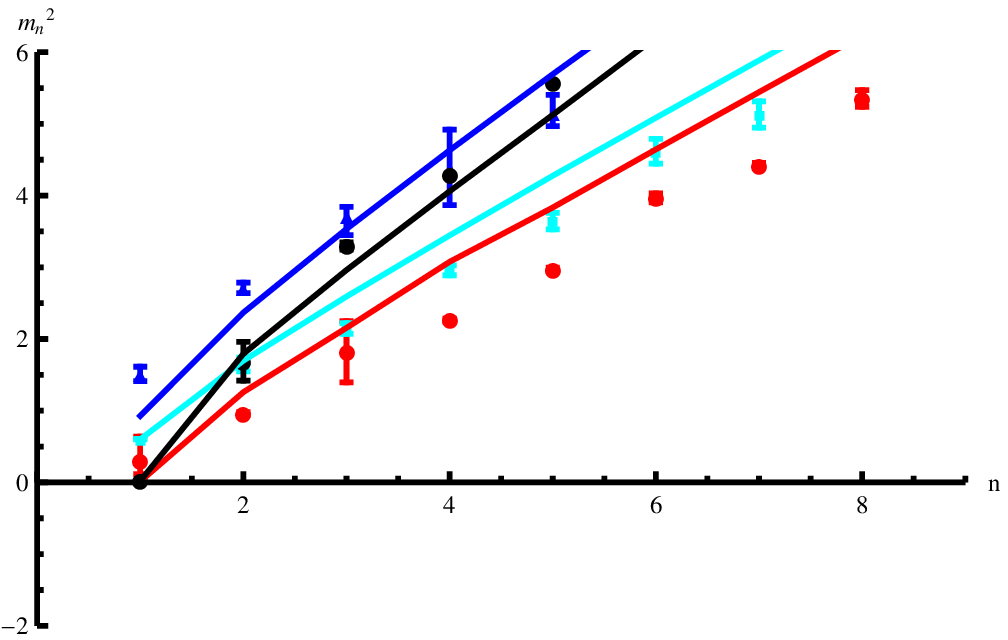} \hspace*{0.1cm}
\epsfxsize=6.5 cm \epsfysize=6.5 cm \epsfbox{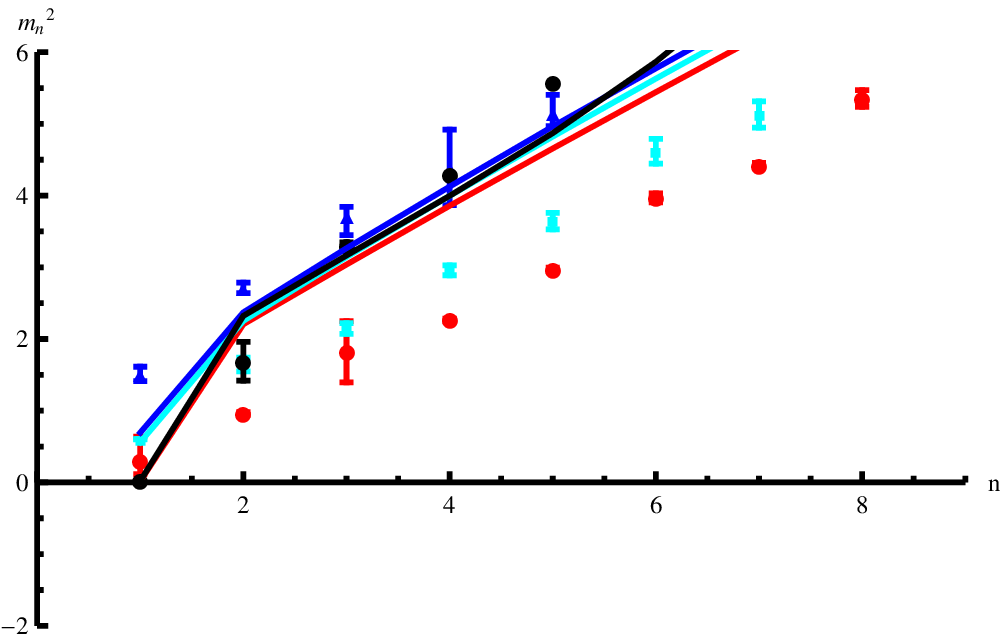} \vskip -0.05cm
\hskip 0.15 cm
\textbf{( case-a ) } \hskip 6.5 cm \textbf{(case-b )} \\
\end{center}
\caption[]{Meson spectra in the SWXY model comparing with experimental data
in Table \ref{exp-mass}. The parameters $\mu=445 {\rm MeV}$ (their $\mu_d$), and
$m_q=4.98 {\rm MeV}$, $\sigma=(255 {\rm MeV})^3$ are used for case-a, and
$m_q=4.25 {\rm MeV}$, $\sigma=(268 {\rm MeV})^3$ are used for case-b.}
\label{all-SW-ab}
\end{figure}

The meson spectra for the SWXY model are shown in Fig.\ref{all-SW-ab}(we only take the Model III in their original paper as an example), where
for case-a(their IIIa) they have used parameters as $\mu=445 {\rm MeV}, m_q=4.98 {\rm MeV},
\sigma=(255 {\rm MeV})^3$, and for case-b(their IIIb), they have use parameters as
$\mu=445 {\rm MeV}, m_q=4.25 {\rm MeV},\sigma=(268 {\rm MeV})^3$.  It is found that
for case-a, the Regge slopes for the scalar and vector meson spectra are the same,
and the Regge slopes for the pseudo-scalar and axial-vector meson spectra are the same, while the slopes of the two groups are different.
For case-b, all spectra are degenerate.

It is worthy of mentioning that the meson spectra are compared with experimental data taken in
Table \ref{exp-mass}, which are the same as in Ref. \cite{Gherghetta-Kapusta-Kelley},
and different from the experimental data taken in Ref. \cite{YLWu}. As for which data
should be taken properly, and whether there should be chiral symmetry restoration at
high excitation states \cite{Huang:2007fv,chiralr,chiralb}, we leave them
as open questions.

\section{Two flavor system: the dynamical soft-wall model}
\label{sec-hadronspectra-dhQCD}

A successful holographic QCD model should describe chiral symmetry breaking,
and at the same time should describe both the Regge trajectories of hadron
spectra and linear quark potential, two aspects in the manifestation of
color confinement. Thus how to naturally incorporate all these important features
into a single AdS/QCD model and obtain the consistent mass spectra remains a challenging
and interesting task. In this section, we provide a fully dynamical soft-wall holographic
QCD model formulated in the graviton-dilaton-scalar system, which can incorporate chiral
symmetry breaking, Regge spectra as well as linear quark potential.

\subsection{Dynamical soft-wall model: the graviton-dilaton-scalar system}

As we have shown in Sec. II that the pure gluodynamics can be described very well
by the quenched dynamical soft-wall model formulated in the graviton-dilaton system.
The quadratic correction of dilaton background at IR related to the gluon condensate
in the vacuum can produce the linear confinement, including linear Regge spectra and
the linear heavy quark potential.
We now add light flavors in terms of meson fields on the gluodynamical background. The total
5D action for the graviton-dilaton-scalar system takes the following form:
\begin{eqnarray}
 S=S_G + \frac{N_f}{N_c} S_{KKSS},
 \label{fullaction}
\end{eqnarray}
with
\begin{eqnarray}
 S_G=&&\frac{1}{16\pi G_5}\int
 d^5x\sqrt{g_s}e^{-2\Phi}\big(R+4\partial_M\Phi\partial^M\Phi-V_G(\Phi)\big), \\
 S_{KKSS}=&&-\int d^5x
 \sqrt{g_s}e^{-\Phi}Tr(|DX|^2+V_X(X^+X, \Phi)+\frac{1}{4g_5^2}(F_L^2+F_R^2)).
\end{eqnarray}
It is noticed that $S_G$ is the 5D action for gluons in terms of dilaton field $\Phi$ and
takes the same form as Eq.(\ref{action-graviton-dilaton}), here we have assumed the action
is in the string frame. $S_{KKSS}$ is the 5D action for mesons propagating on the
dilaton background and takes the same form as the general KKSS action Eq.(\ref{action-KKSS}).
$V_G(\Phi)$ and $V_X(X^+X,\Phi)$ are potentials for dilaton field and scalar field, respectively.
It is noticed that the scalar field might mix with the gluon fields, therefore we have chosen
a general form for the scalar potential $V_X(X^+X,\Phi)$.

In the vacuum, it is assumed that there are both gluon condensate and chiral condensate.
The dilaton background field $\Phi$ is supposed to be dual to some kind of gluodynamics
in QCD vacuum. For the pure gluon system, we have shown in Sec. \ref{sec-glueball}, that
two forms of quadratic correction to the dilaton background field at IR can produce
glueball spectra in agreement with lattice data. In the following we define two types
of graviton-dilaton-scalar models corresponding to two different forms of
dilaton background field:
\begin{eqnarray}
Dilaton ~in ~Mod~I: ~~ & & \Phi(z)=\mu_G^2z^2 \label{mod1} \\
Dilaton~in ~Mod~II: ~~ & & \Phi(z)=\mu_G^2z^2\tanh(\mu_{G^2}^4z^2/\mu_G^2).
\label{mod2}
\end{eqnarray}

With the quadratic dilaton background field, Mod I can be regarded as a selfconsistent
KKSS model, where the metric structure is not AdS$_5$ anymore but automatically deformed
at IR. As discussed previously, with quadratic dilaton background field, we may encounter
the gauge invariant problem for the dimension-2 gluon operator. To avoid the gauge
non-invariant problem and to meet the requirement of gauge/gravity duality,
we take the dilaton field with quartic form at UV and quadratic form at IR as in
Eq.(\ref{mod2}).

\subsection{Background fields in the vacuum with chiral and gluon condensate}

The scalar field $X(z)$ is a complex field as shown in Eq.(\ref{scalarfield}) and
it is expected that the scalar field takes a nonzero vacuum expectation value (VEV)
$\chi(z)$.

It's easy to get the 5D action for the vacuum background:
\begin{equation}
S_{vac} = S_{G,vac}+ \frac{N_f}{N_c} S_{KKSS,vac},
\end{equation}
with
\begin{eqnarray}
S_{G,vac} &=&\frac{1}{16\pi G_5}\int d^5x\sqrt{g_s}e^{-2\Phi}\big(R
  +4\partial_M\Phi\partial^M\Phi-V_G(\Phi)\big) \\
 S_{KKSS,vac}&=& - \int d^5x \sqrt{g_s}e^{-\Phi}(\frac{1}{2}\partial_M\chi
 \partial^M \chi+V_{C}(\chi,\Phi))
\end{eqnarray}
where we have defined $V_C=  Tr(V_X)$
For further convenience, we define
\begin{equation}
V_{C,\chi}=\frac{\partial V_C}{\partial \chi}, ~~
V_{C,\chi\chi}=\frac{\partial^2V_C}{\partial^2\chi}.
\end{equation}
By Redefinition: $ L^{\frac{3}{2}}\chi\rightarrow
\chi,L^3V_C\rightarrow V_C, \frac{16\pi
G_5 N_f}{L^3 N_c}\rightarrow \lambda $ we can set all the fields
and constants to be dimensionless, and the vacuum action takes the form of
\begin{eqnarray}
S_{vac}=\frac{1}{16 \pi G_5} \int d^5 x &&\sqrt{g_s} \big\{ e^{-2\Phi}[R
+4\partial_M\Phi \partial^M \Phi - V_G(\Phi)] \nonumber\\
&& - \lambda e^{-\Phi} (\frac{1}{2} \partial_M\chi \partial ^M \chi
+ V_C(\chi,\Phi))\big\}
\end{eqnarray}

After the frame transformation $g^s_{MN}=g^E_{MN}e^{\frac{2}{3}\Phi}$, the action
$S_{vac}$ in the Einstein frame takes the following form
\begin{eqnarray}
S_{vac}=\frac{1}{16 \pi G_5} \int d^5 x &&\sqrt{g_E} \big\{[R_E-\frac{4}{3}\partial_M\Phi
\partial^M \Phi - V_G^E(\Phi)] \nonumber\\
&& - \lambda e^{\Phi}( \frac{1}{2} \partial_M\chi \partial ^M \chi
+ e^{\frac{4}{3}\Phi} V_C(\chi,\Phi))\big\}.
\end{eqnarray}
The Einstein equation and field equations in the Einstein frame have the expression of
\begin{eqnarray}
 E_{MN}+\frac{1}{2}g^E_{MN}\left(\frac{4}{3}\partial_l\Phi\partial^l\Phi
  +V_G^E(\Phi)+\lambda(\frac{1}{2}e^{\Phi}\partial_l\chi\partial^l\chi
  +e^{\frac{7}{3}\Phi}V_C(\chi,\Phi))\right)& & \\
 -\frac{4}{3}\partial_M\Phi\partial_N\Phi-\frac{\lambda}{2}
 e^{\Phi}\partial_M\chi\partial_N\chi &=&0, \\
 \frac{8}{3\sqrt{g_E}}\partial_M(\sqrt{g_E}\partial^M\Phi)
 -\lambda\frac{1}{2}e^{\Phi}\partial_M\chi\partial^M\chi
 -\partial_{\Phi}\left(V_G(\Phi)+\lambda e^{\frac{7}{3}\Phi}V_C(\chi,\Phi)\right)&=&0,\\
 \lambda\frac{1}{\sqrt{g_E}}\partial_M(\sqrt{g_E}e^{\Phi}\partial^M\chi)
 -\partial_{\chi}\left(V_G(\Phi)+\lambda
 e^{\frac{7}{3}\Phi}V_C(\chi,\Phi)\right)&=&0.
\end{eqnarray}
We can derive the three coupled field equations in the string frame as
\begin{eqnarray}
 -A_s^{''}+A_s^{'2}+\frac{2}{3}\Phi^{''}-\frac{4}{3}A_s^{'}\Phi^{'}
 -\frac{\lambda}{6}e^{\Phi}\chi^{'2}&=&0, \label{Eq-As-Phi} \\
 \Phi^{''}+(3A_s^{'}-2\Phi^{'})\Phi^{'}-\frac{3\lambda}{16}e^{\Phi}\chi^{'2}
 -\frac{3}{8}e^{2A_s-\frac{4}{3}\Phi}\partial_{\Phi}\left(V_G(\Phi)
 +\lambda e^{\frac{7}{3}\Phi}V_C(\chi,\Phi)\right)&=&0, \label{Eq-VG}\\
 \chi^{''}+(3A_s^{'}-\Phi^{'})\chi^{'}-e^{2A_s}V_{C,\chi}(\chi,\Phi)&=&0. \label{Eq-Vc}
\end{eqnarray}
If we know the form of the dilaton field $\Phi$ and the scalar field $\chi$,
then the metric $A_s$, the dilaton potential $V_G(\Phi)$ and the scalar potential
$V_C(\chi,\Phi)$ should be self-consistently solved from the above three coupled equations.

\subsection{Chiral symmetry breaking and linear confinement}

For "Mod~I" with positive quadratic dilaton background $\Phi(z)=\mu_G^2 z^2$,
we will constrain the form of scalar VEV from the linear confinement.

\vskip 0.2cm
{\bf\it The UV asymptotic form of $\chi(z)$}
\vskip 0.2cm

As proposed in the \cite{Cherman:2008eh}, at the ultraviolet(UV) region, the scalar field takes the
following asymptotic form,
\begin{eqnarray}
\chi(z) \stackrel{z \rightarrow 0}{\longrightarrow} m_q \zeta z+\frac{\sigma}{\zeta} z^3,
\label{chi-IR}
\end{eqnarray}
where $m_q$ is the current quark mass, and $\sigma$ is the quark
antiquark condensate, and $\zeta$ is a normalization constant and is
fixed as $\zeta^2=\frac{N_c^2}{4\pi^2N_f}$ with $N_c=3, N_f=2$.

\vskip 0.2cm
{\bf\it The IR asymptotic form of $\chi(z)$ constrained from linear quark potential}
\vskip 0.2cm

The linear behavior of quark-antiquark static potential
in the heavy quark mass limit $m_Q\rightarrow \infty$ can describe
the permanent confinement property of QCD. In Sec. \ref{sec-G-HQ},
we have derived the heavy quark potential under the general metric background
$A_s$, and we have observed that for the metric structure $b_s=e^{A_s}$,
 if there exists a point $z_c$ where $b_s^{'}(z_c)\rightarrow 0$,
then one can extract the string tension $\sigma_s$ of the linear potential as
\begin{eqnarray}
\sigma_s=\frac{V_{\bar{q}q}(z_0)}{R_{\bar{q}q}(z_0)}\overset{z_0\rightarrow z_c}{\longrightarrow}
 \frac{g_p}{2\pi} b_s^2(z_c). \label{stringtension}
\end{eqnarray}
Therefore, the necessary condition for the linear part of the
$Q-\bar{Q}$ potential is that there exists one point $z_c$ or one region,
where $b_s^{'}(z)\rightarrow 0,z\rightarrow z_c$ while $b_s(z)$ keeps
finite. For simplicity, we can take the following constraint on the
metric structure at IR(taking $z_c=\infty$):
\begin{equation}
A_s^{'}(z) \stackrel{z \rightarrow \infty}{\longrightarrow} 0,
A_s(z) \stackrel{z \rightarrow \infty}{\longrightarrow} {\rm Const}.
\label{Metric-IR}
\end{equation}
Under the condition of Eq.(\ref{Metric-IR}),
the equation of motion Eq.(\ref{Eq-As-Phi}) in the IR takes the
following simple form:
\begin{equation}
\frac{2}{3}\Phi^{''}-\frac{\lambda}{6}e^{\Phi}\chi^{'2}=0,
\label{phi-chi-IR}
\end{equation}
which provides a relation between the chiral condensate and
low energy gluodynamics at IR. The asymptotic form of
$\chi(z)$ at IR can be solved as:
\begin{equation}
\chi(z)\stackrel{z \rightarrow \infty}{\longrightarrow} \sqrt{8/\lambda}\mu_G e^{-\Phi/2}.
\label{chi-UV}
\end{equation}

\begin{table}
\begin{center}
\begin{tabular}{cccccccc}
\hline\hline
  ~                     &      Mod~IA        & Mod~IB      &Mod~IIA    &Mod~IIB    \\   \hline
 $G_5/L^3$              &      $0.75$        & $0.75$      & $0.75$    &$0.75$     \\
 $m_q$~(MeV)            &      $5.8$         & $5.0$       &  $8.4$    &$6.2$      \\
$\sigma^{1/3}~(MeV)$    &      $180$         &$240$       &  $165$     &$226$       \\
$\mu_G$                 &      0.43           & 0.43      & 0.43       & 0.43       \\
$\mu_{G^2}$             &      -             & -          & 0.43       & 0.43        \\
\hline\hline
\end{tabular}
\caption{Two sets of parameters for both Mod I and Mod II.}
\label{parameters}
\end{center}
\end{table}

\begin{figure}[h]
\begin{center}
\epsfxsize=5.5 cm \epsfysize=5.5 cm \epsfbox{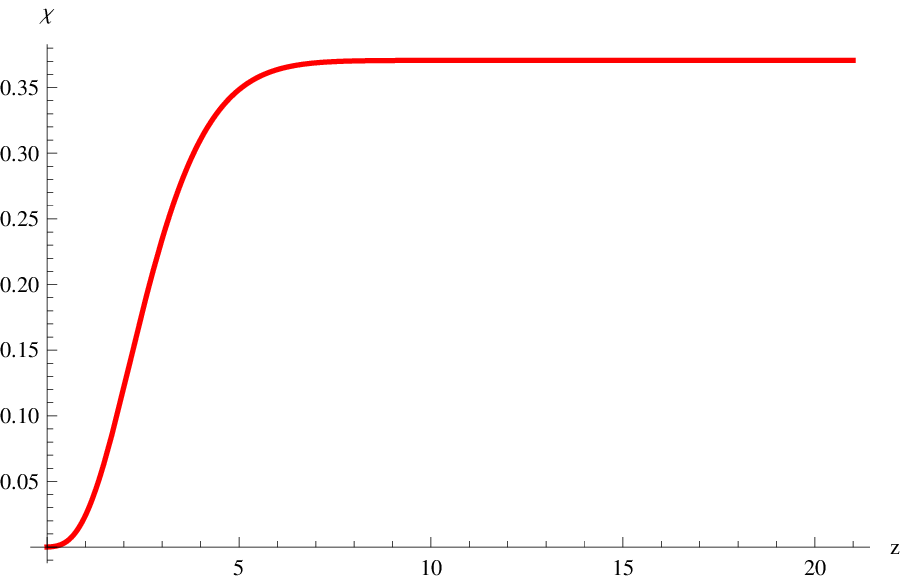} \hspace*{0.2cm}
\epsfxsize=5.5 cm \epsfysize=5.5 cm \epsfbox{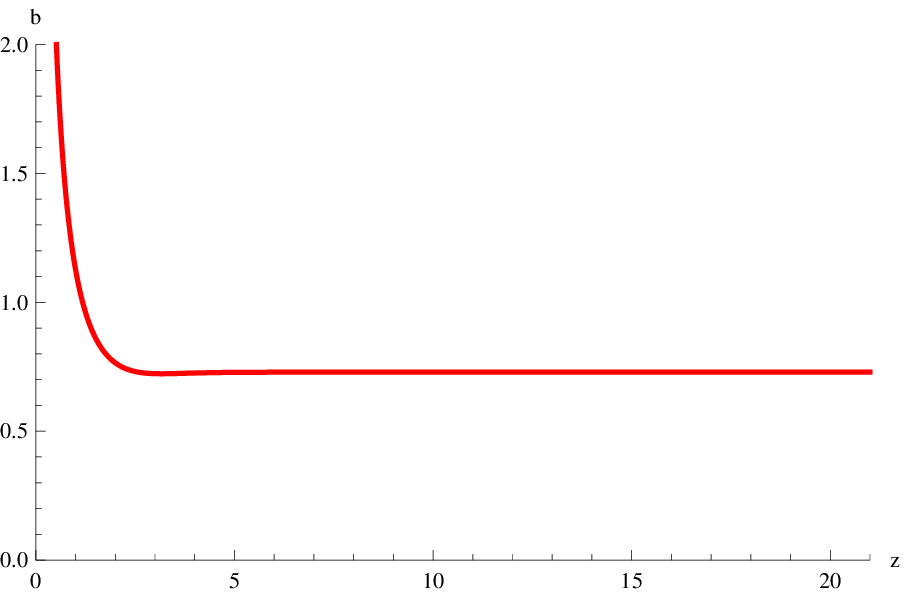} \vskip -0.005cm
\hskip 0.15 cm
\textbf{( $IA:\chi$ ) } \hskip 6.5 cm \textbf{( $IA:b_s$ )} \\
\epsfxsize=5.5 cm \epsfysize=5.5 cm \epsfbox{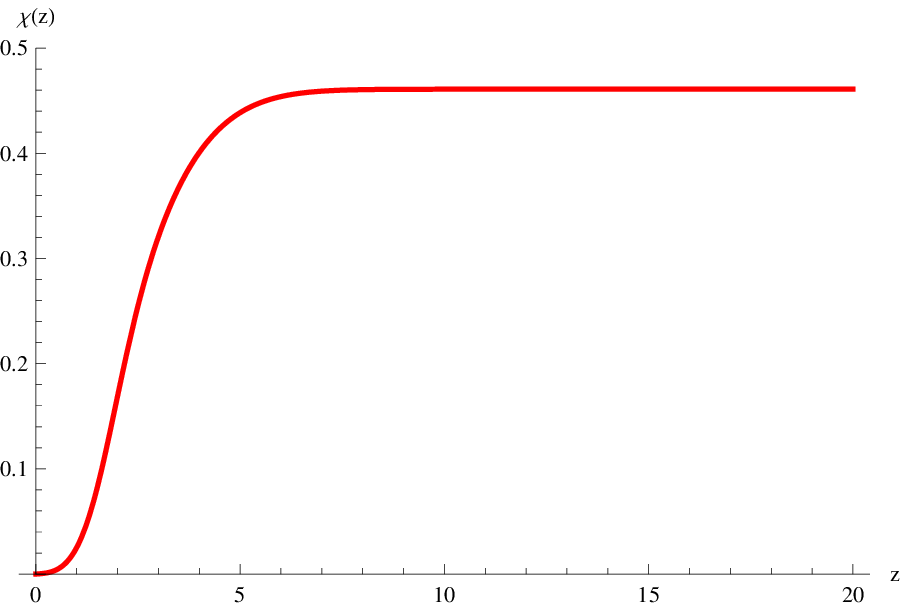} \hspace*{0.2cm}
\epsfxsize=5.5 cm \epsfysize=5.5 cm \epsfbox{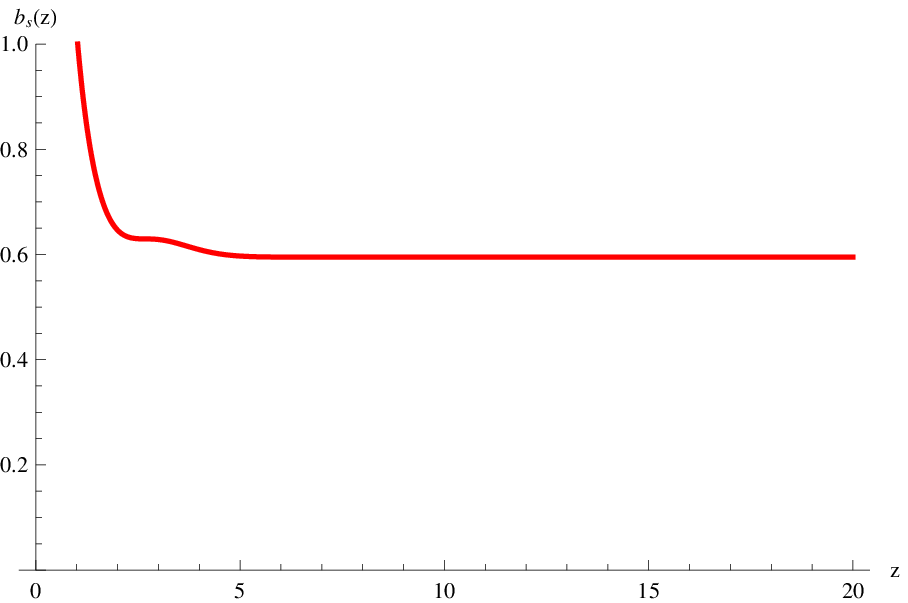} \vskip -0.005cm
\hskip 0.15 cm
\textbf{( $IIA:\chi$ ) } \hskip 6.5 cm \textbf{( $IIA:b_s$ )} \\
\end{center}
\caption{Scalar VEV $\chi(z)$ and solved metric structure $b_s$ as functions of $z$
for Mod IA and Mod IIA. }
\label{chi-bs}
\end{figure}

\vskip 0.2cm
{\bf\it The full form of $\chi(z)$}
\vskip 0.2cm

To match the asymptotic forms both at UV and IR in Eqs.(\ref{chi-IR}) and (\ref{chi-UV}),
for the dilaton field Eq.(\ref{mod1}),$\chi$ can be parameterized as
\begin{eqnarray}\label{chiz}
\chi^{'}(z)=\sqrt{8/\lambda}\mu_G e^{-\Phi/2}(1+c_1 e^{- \Phi}+c_2
e^{-2\Phi}),
\end{eqnarray}
of which with the exponential suppressing Eq.(\ref{chi-UV}) is satisfied and by taking
$c_1=-2+\frac{5\sqrt{2\lambda}m_q\zeta}{8\mu_G}+\frac{3\sqrt{2\lambda}\sigma}{4\zeta
\mu_G^3},c_2=1-\frac{3\sqrt{2\lambda}m_q\zeta}{8\mu_G}-\frac{3\sqrt{2\lambda}\sigma}{4\zeta
\mu_G^3}$ Eq.(\ref{chi-IR}) is satisfied. Solving Eq.(\ref{chiz}), we can obtain the full expression for the scalar VEV,
which takes the following form:
\begin{eqnarray}
 \chi(z)=&&\frac{1}{30 \zeta  \mu_G ^3}\sqrt{\frac{\pi }{2\lambda}}
 \big(5 \sqrt{3} \text{Erf}\left(\sqrt{\frac{3}{2}} \mu_G  z\right)
 \left(-8 \sqrt{2} \zeta  \mu_G ^3+6 \sqrt{\lambda} \sigma
 +5 \zeta ^2 \sqrt{\lambda} \mu_G ^2 m_q \right)\nonumber\\
 &&+3 \big(\sqrt{5} \text{Erf}\left(\sqrt{\frac{5}{2}} \mu_G  z\right)
 \left(4 \sqrt{2} \zeta  \mu_G ^3-6 \sqrt{\lambda} \sigma -3 \zeta ^2 \sqrt{\lambda}
 \mu_G ^2 m_q\right) \nonumber \\
 && +20 \sqrt{2} \zeta  \mu_G ^3 \text{Erf}
 \left(\frac{\mu_G  z}{\sqrt{2}}\right)\big)\big),
 \label{sol-chi}
\end{eqnarray}
where $g_5^2=4\pi^2\frac{N_f}{N_c}$ and $\zeta^2=\frac{N_c^2}{4\pi^2 N_f}$.

\begin{figure}[h]
\begin{center}
\epsfxsize=6.5 cm \epsfysize=6.5 cm \epsfbox{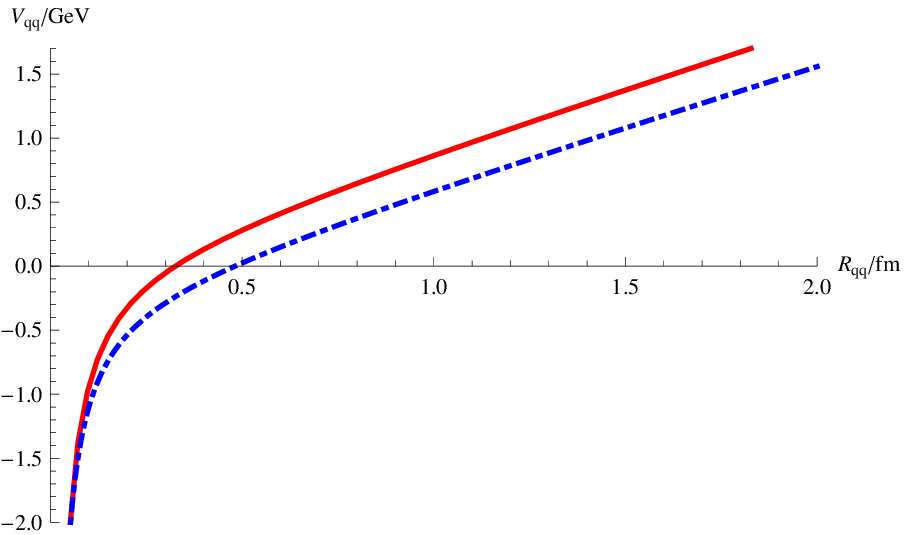}\hspace*{0.2cm}
\epsfxsize=6.5 cm \epsfysize=6.5 cm \epsfbox{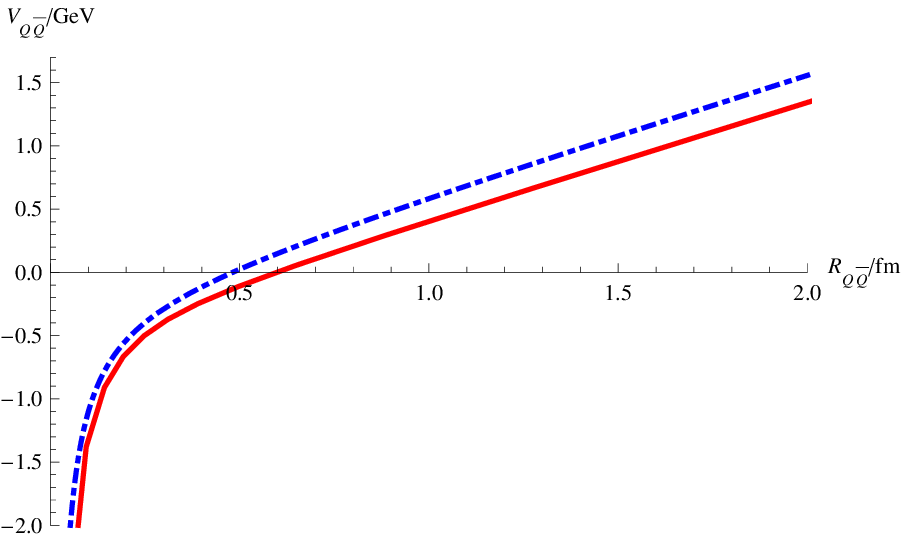}\vskip -0.005cm
\hskip 0.15 cm
\textbf{( $IA:V_{Q\bar{Q}}$ ) } \hskip 6.5 cm \textbf{( $IIA:V_{Q\bar{Q}}$ )} \\
\end{center}
\caption{Heavy quark potential $V_{Q{\bar Q}}$ as a function of $R_{Q{\bar Q}}$ for
 Mod IA (with $g_p=2.2$) and Mod IIA (with $g_p=2.8$) are shown in solid lines compared
with the Cornell potential shown in dot-dashed lines. } \label{Vqq-full}
\end{figure}

Similarly, for "Mod II" with the the dilaton field Eq.(\ref{mod2}), $\chi$ can be
parameterized as
\begin{eqnarray}
\chi^{'}(z)=\sqrt{8/\lambda}\mu_G e^{-\Phi/2}(1+d_1 e^{- \Phi}+d_2 z^2
e^{-2\Phi}-\frac{1}{2}e^{-3\Phi}).
\label{sol-chi2}
\end{eqnarray}
To satisfy Eq.(\ref{chi-IR}) one needs $d_1=-\frac{1}{2}+\frac{\sqrt{\lambda}m_q \zeta}{2\sqrt{2}\mu_G}, d_2=\frac{3\sqrt{\lambda}\sigma}{2\sqrt{2}\zeta\mu_G}$.

In our following numerical calculations, we will use two sets of parameters for each model,
i.e. we take Mod IA, Mod IB, Mod IIA and Mod IIB and the corresponding parameters are given
in Table \ref{parameters}. In order to fit the Regge slope of meson spectra, $\mu_G$ is fixed
as $0.43 {\rm GeV}$ which is the same as in the KKSS model, in our parameterization, as long as
$\mu_{G^2}>\mu_G$, the results for meson spectra are not sensitive to the value of $\mu_{G^2}$.
So we take $\mu_{G^2}=\mu_G$ in "Mod IIA" and "Mod IIB". As we will show later, these four sets
of parameters can produce meson spectra in good agreement with experimental data. With parameters in set A, one can produce better result for pion form factor with the price of lower pion decay
constant. With parameters in set B, one can produce better result for pion decay constant, but
worse pion form factor.

With the input of dilaton field $\Phi(z)$ given in Eqs.(\ref{mod1}) and (\ref{mod2}),
and $\chi(z)$ given in Eqs.(\ref{sol-chi}) and (\ref{sol-chi2}),
one can solve the metric $A_s$ or $b_s$ from the equation of motion Eq. (\ref{Eq-As-Phi}).
By taking the parameters in set A for Mod I and Mod II, we show
the numerical results for the scalar VEV $\chi(z)$ and the solved metric structure $b_s(z)$
in Fig.\ref{chi-bs}. It is found that both $\chi(z)$ and $b_s(z)$ are saturate at IR.

The heavy quark potentials under the solved metric structure for Mod IA and Mod IIA
are also shown in Fig. \ref{Vqq-full} by the solid lines and comparing with the Cornell
potential $V^{Cornell}(R)=-\frac{\kappa}{R}+\sigma_{s}R+V_0$  with $\kappa\approx
0.48$, $\sigma_{s}\approx 0.183 {\rm GeV}^{2}$ and $V_0=-0.25 {\rm
GeV}$. It is observed that the heavy quark potential produced in our model
including the back-reaction from light flavor dynamics agree well with the
Cornell potential.

\subsection{Meson spectra in the graviton-dilaton-scalar system}
\label{meson-quadratic-quatic}

With the dilaton background field $\Phi(z)$ in Eqs.(\ref{mod1}) and (\ref{mod2}),
and the scalar background field $\chi(z)$ given in Eqs.(\ref{sol-chi}) and (\ref{sol-chi2}),
we have solved the metric $A_s$ or $b_s$ from
the equation of motion  Eq. (\ref{Eq-As-Phi}). Now we are ready to derive the meson spectra
in the dynamical soft-wall model.

\subsubsection{Scalar spectra}

\begin{table}
\begin{center}
\begin{tabular}{cccccccc}
\hline\hline
        n & $f_0$~Exp~(MeV)        & Mod~IA~(MeV)   & Mod~IB~(MeV)     &Mod~IIA~(MeV)      &Mod~IIB~(MeV) \\
\hline
        1 & $550^{+250}_{-150}$    &421             & 231              &580                &187               \\
        2 & $980 \pm 10$           &1043            & 1106             &1066               &1078               \\
        3 & $1350 \pm 150$         &1370            & 1395             &1400               &1434              \\
        4 & $1505 \pm 6$           &1625            & 1632             &1656               &1685               \\
        5 & $1873 \pm 7$           &1843            & 1846             &1873               &1890              \\
        6 & $1992 \pm 16$          &2036            & 2039             &2064               &2068               \\
        7 & $2103 \pm 8$           &2212            & 2215             &2237               &2234               \\
        8 &  $2314 \pm 25$         &2375            & 2376             &2396               &2392               \\
\hline\hline
\end{tabular}
\caption{The experimental and predicted mass spectra for scalar
mesons $f_0$.} \label{scalarmasses}
\end{center}
\end{table}

\begin{figure}[h]
\begin{center}
\epsfxsize=6.5 cm \epsfysize=6.5 cm \epsfbox{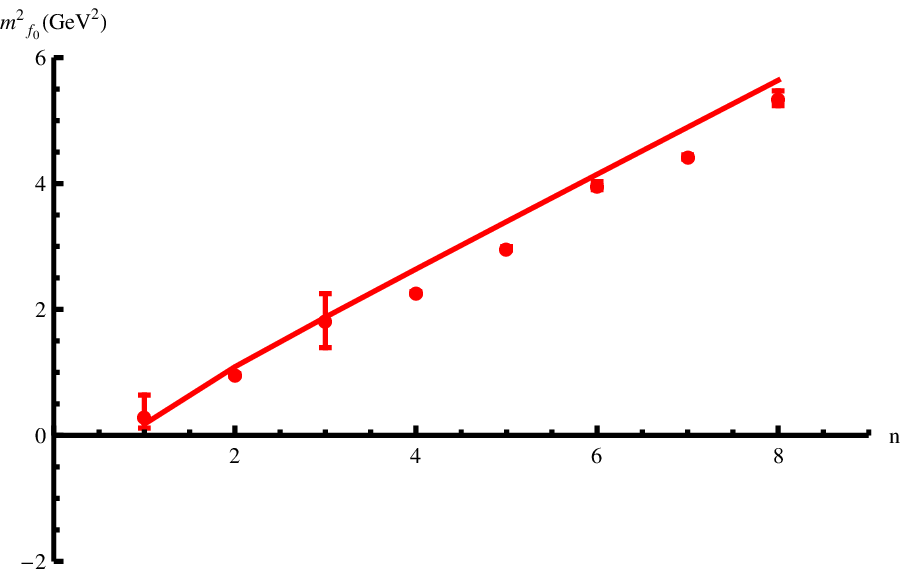} \hspace*{0.1cm}
\epsfxsize=6.5 cm \epsfysize=6.5 cm \epsfbox{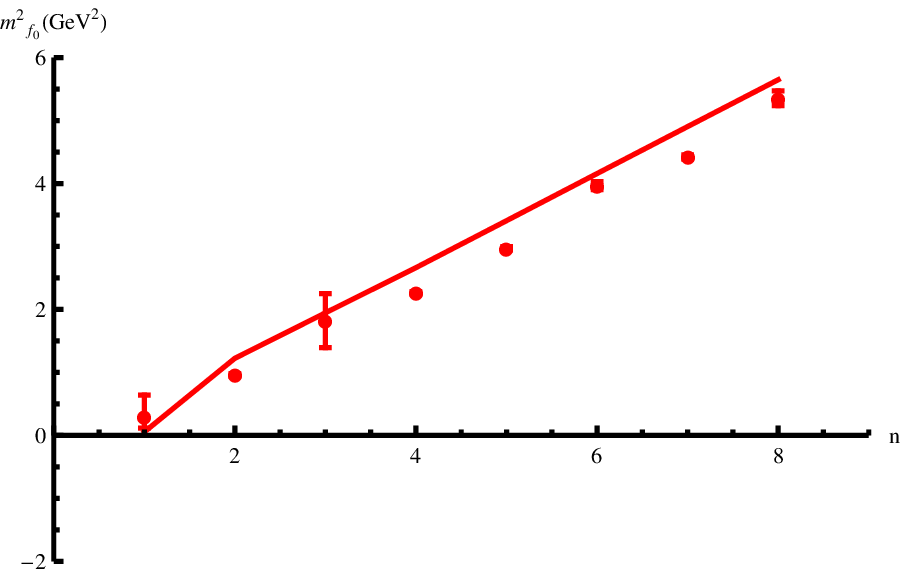} \vskip -0.05cm
\hskip 0.15 cm
\textbf{( Mod IA ) } \hskip 6.5 cm \textbf{( Mod IB )} \\
\epsfxsize=6.5 cm \epsfysize=6.5 cm \epsfbox{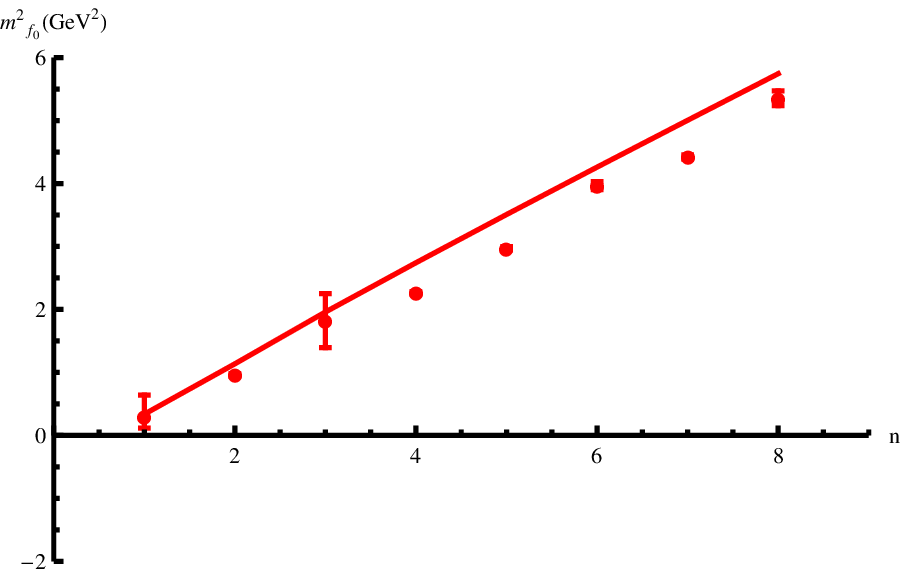} \hspace*{0.1cm}
\epsfxsize=6.5 cm \epsfysize=6.5 cm \epsfbox{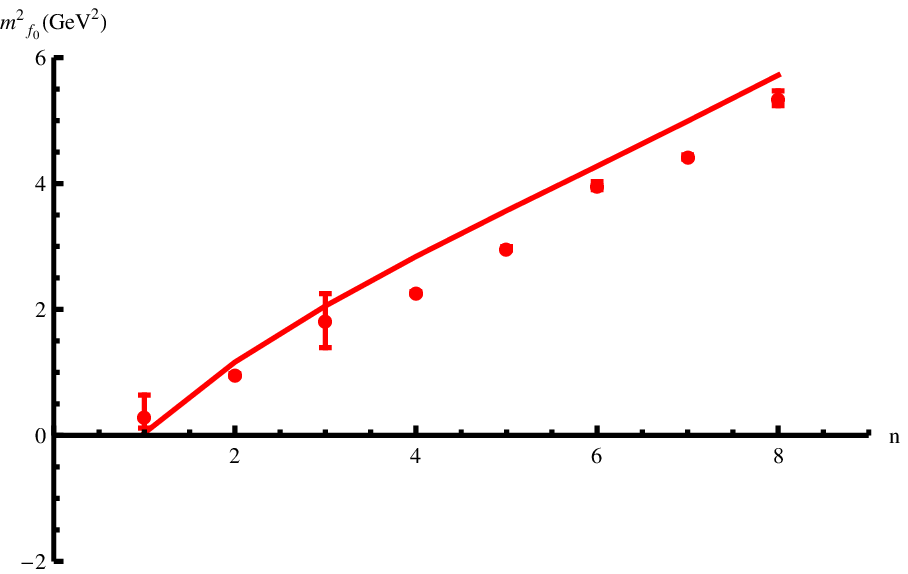} \vskip -0.05cm
\hskip 0.15 cm
\textbf{( Mod IIA ) } \hskip 6.5 cm \textbf{( Mod IIB )} \\
\end{center}
\caption[]{Scalar meson spectra $m_{f_0,n}^2$ as functions of $n$ for Mod I and II
defined in Table \ref{parameters}.} \label{scalarmassespic}
\end{figure}

\begin{figure}[h]
\begin{center}
\epsfxsize=10.0 cm \epsfysize=7.5 cm \epsfbox{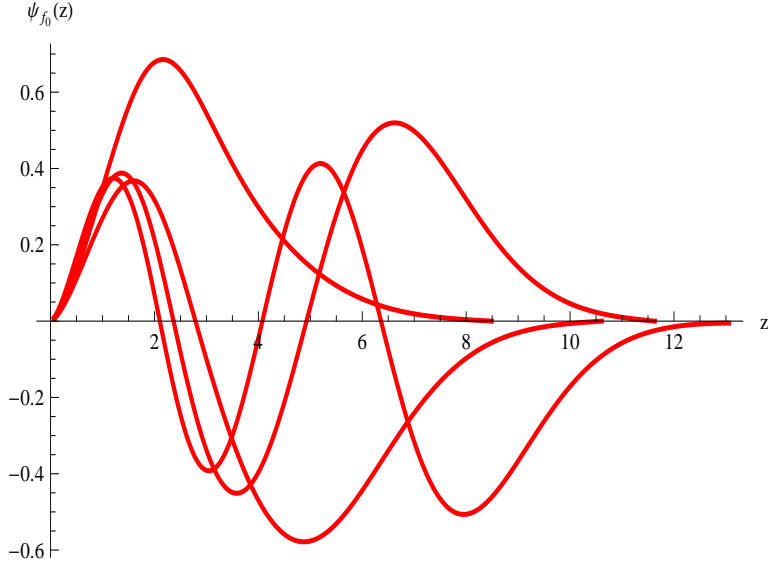}
\end{center}
\caption{The scalar wave function $\psi_{f_0,n}(z)$ as function of $z$ for Mod IB. }
\label{f0wavepic}
\end{figure}

The action of the scalar perturbation is
\begin{eqnarray}
 S_{s}= -2\frac{N_f}{N_cL^3}\int d^5x
 e^{-\Phi}\sqrt{g_s}(\partial_zS\partial^zS+\partial_\mu S\partial^\mu
 S+V_{C,\chi\chi}(\chi,\Phi)S^2),
\end{eqnarray}
and the equation of motion for the scalar perturbation after doing
the $KK$ modes expansion is
\begin{eqnarray}
-e^{-(3A_s-\Phi)}\partial_z(e^{3A_s-\Phi}\partial_zs_n)+
e^{2A_s}V_{C,\chi\chi}(\chi,\Phi)s_n=m_n^2s_n.
\end{eqnarray}
By doing the transformation $s_n\rightarrow s_n
e^{-(3A_s-\Phi)/2}$, one can get the schrodinger like
equation
\begin{eqnarray}\label{scalar-sn}
-s_n^{''}+V_s(z)s_n=m_n^2s_n
\end{eqnarray}
with the schrodinger potential
\begin{eqnarray}
V_s(z)=\frac{3A_s^{''}-\Phi^{''}}{2}+\frac{(3A_s^{'}-\Phi^{'})^2}{4}
+e^{2A_s}V_{C,\chi\chi}(\chi,\Phi).
\end{eqnarray}
Assuming the scalar potential can be separated into
\begin{equation}
V_{C}(\chi,\Phi)=e^{f(\Phi)}V_{c}(\chi),
\end{equation}
we have
\begin{eqnarray}
&&e^{2A_s}V_{C,\chi\chi}=e^{2A_s+f(\Phi)}V_{c,\chi\chi}
=e^{2A_s+f(\Phi)}\frac{\partial_z(V_{c,\chi})}{\chi^{'}}\nonumber\\
\end{eqnarray}
By using Eq.(\ref{Eq-Vc}), we can have
\begin{eqnarray}
e^{2A_s}V_{C,\chi\chi}
&&=e^{2A_s+f(\Phi)}\frac{\partial_z(e^{-(2A_s+f(\Phi))}
(\chi^{''}+(3A_s^{'}-\Phi^{'})\chi^{'}))}{\chi^{'}}\nonumber\\
&&=\frac{\chi^{'''}}{\chi^{'}}+(A_s^{'}-\Phi^{'}
-f_{,\Phi}\Phi^{'})\frac{\chi^{''}}{\chi^{'}}
+3A_s^{''}-\Phi^{''} \nonumber \\
&& -(2A_s^{'}+f_{,\Phi}\Phi^{'})(3A_s^{'}-\Phi^{'}).
\end{eqnarray}

In our solution, $\chi^{'}(z) \propto e^{-\frac{1}{2}\Phi}$, so the leading term of
$e^{2A_s}V_{C,\chi\chi}$ is $\frac{\Phi^{'2}}{4}-\frac{\Phi^{'2}}{2}(-1-f_{,\Phi})
+f_{,\Phi}\Phi^{'2}=(\frac{3}{4}+\frac{3 f_{,\Phi}}{2})\Phi^{'2}$.
The Regge behavior of the spectral is determined by the leading IR behavior,
the coefficient before $\Phi^{'2}$ is proportional to the Regge slope.
In order to be consistent with the experimental data(the universal Regge slope in different sectors), we
need $\frac{3}{4}+\frac{3 f_{,\Phi}}{2}\rightarrow0$ in the IR region,
so the leading term of $f_{,\Phi}=-\frac{1}{2}$ and
$f(\Phi)\rightarrow-\frac{1}{2}\Phi$ in large $\Phi$ region. Then we examine the IR behavior
of the EOM of $\chi$,
\begin{eqnarray}
-\frac{\Phi^{'}}{2}e^{-\frac{\Phi}{2}}-\Phi^{'}e^{-\frac{\Phi}{2}}\propto
e^{f(\Phi)}V_{c,\chi}(\chi)
\end{eqnarray}
Note that when $z\rightarrow\infty$, we have $\Phi^{'}\propto z \propto
\sqrt{\Phi}$ and $\chi\rightarrow const, V_{c,\chi}(\chi)\rightarrow const$,
then we can know that when $\Phi\rightarrow\infty$, $f(\Phi)=-\frac{1}{2}\Phi
+\frac{1}{2}\log{\Phi}$. If we hope in the small $\Phi$ region, $e^{f(\Phi)}=1$,
a simplest choice is
\begin{equation}
f(\Phi)=-\frac{\Phi}{2}+\frac{\log(1+\Phi)}{2},
\end{equation}
which leads to the coupling between dilaton background field and the scalar field
at leading order taking the form of
\begin{equation}
V_C(\chi,\Phi) \sim \chi^2 \Phi^2.
\end{equation}

The scalar meson spectra has been numerically calculated with the two sets of parameters
given in Table \ref{parameters} for Mod I and Mod II, respectively.
The predicted scalar meson mass is shown in Table \ref{scalarmasses}, and its
mass square is shown Fig. \ref{scalarmassespic}. The corresponding
wave-functions are shown in Fig. \ref{f0wavepic}. It is observed that for set A
parameters, the produced lowest scalar meson $f_0$ has mass around
$500 {\rm MeV}$ in both Mod I and Mod II, and for set B parameters,
the produced lowest scalar meson $f_0$ has a lower mass around
$200 {\rm MeV}$ in both Mod I and Mod II.

In our graviton-dilaton-scalar system with two different forms of dilaton background,
the lowest scalar state has a positive mass, and the higher excitations behave a Regge
line which agrees well with experimental data.

\subsubsection{Pesudo-Scalar Sector}

The terms of quadratic order in $\pi$ and $\varphi$ ($A^\|_\mu=\partial_\mu \varphi$) is
\begin{eqnarray}
S_{\pi}^{(2)} &=&
-\frac{N_f}{2N_cL^3}\int d^5x
 e^{-\Phi}\sqrt{g_s}(\chi^2\partial_z\pi\partial^z\pi
 +\chi^2\partial_\mu(\pi-\varphi)\partial^\mu(\pi-\varphi) \nonumber \\
 & + & \frac{L^2}{g_5^2}\partial_z\partial_\mu\varphi\partial^z\partial^\mu\varphi).
\end{eqnarray}

\begin{table}
\begin{center}
\begin{tabular}{cccccccc}
\hline\hline
     &  n &$\pi$  Exp~(MeV)      & Mod~IA~(MeV)    &  Mod~IB~(MeV)       &Mod~IIA~(MeV)           &Mod~IIB~(MeV) \\ \hline
      &  1 & $140$               &139.3            & 139.4               &139.6                   &139.1         \\
      &  2 & $1300 \pm 100$      &1343             & 1600                &1505                    &1683              \\
      &  3 & $1816 \pm 14 $      &1755             & 1897                &1832                    &1931                           \\
      &  4 & $2070 $             &2006             & 2116                &2059                    &2138                            \\
      &  5 & $2360 $             &2203             & 2299                &2247                    &2316                             \\
\hline\hline
\end{tabular}
\caption{The experimental and predicted mass spectra for
pseudoscalar mesons $\pi$.} \label{pscalarmasses}
\end{center}
\end{table}

\begin{figure}[h]
\begin{center}
\epsfxsize=6.5 cm \epsfysize=6.5 cm \epsfbox{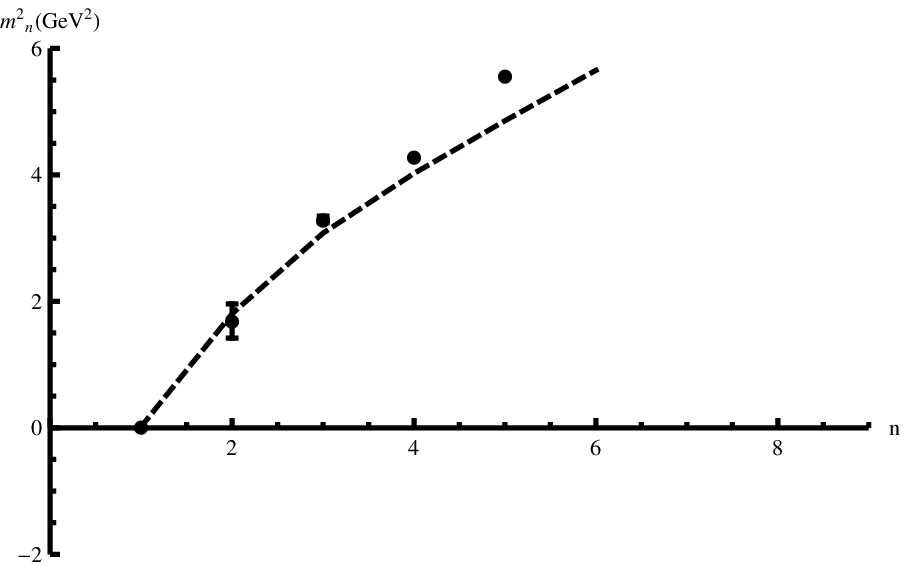} \hspace*{0.1cm}
\epsfxsize=6.5 cm \epsfysize=6.5 cm \epsfbox{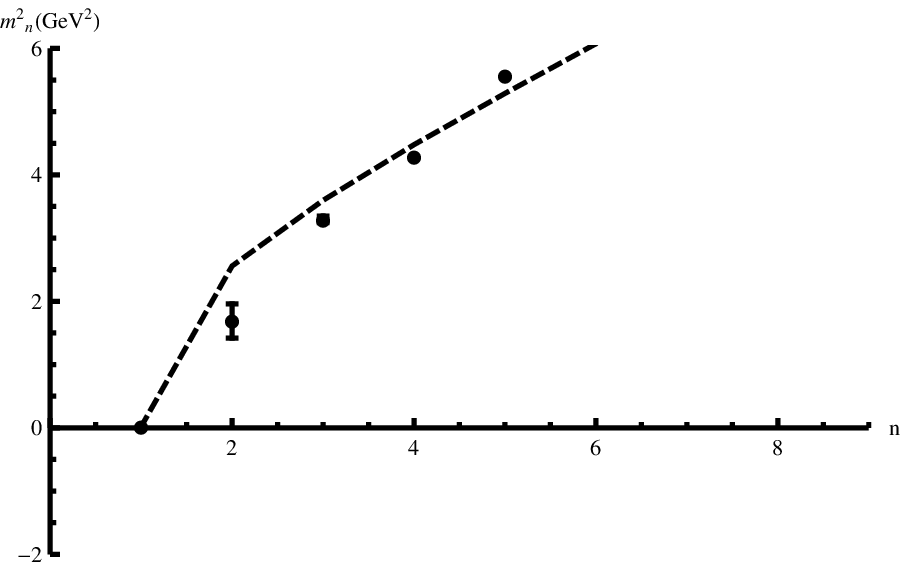} \vskip -0.05cm
\hskip 0.15 cm
\textbf{( Mod IA ) } \hskip 6.5 cm \textbf{( Mod IB )} \\
\epsfxsize=6.5 cm \epsfysize=6.5 cm \epsfbox{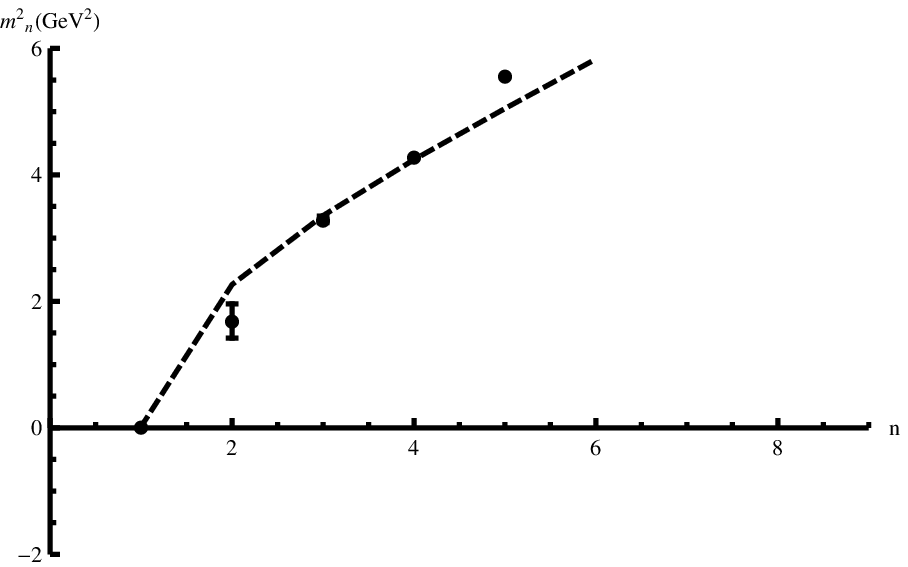} \hspace*{0.1cm}
\epsfxsize=6.5 cm \epsfysize=6.5 cm \epsfbox{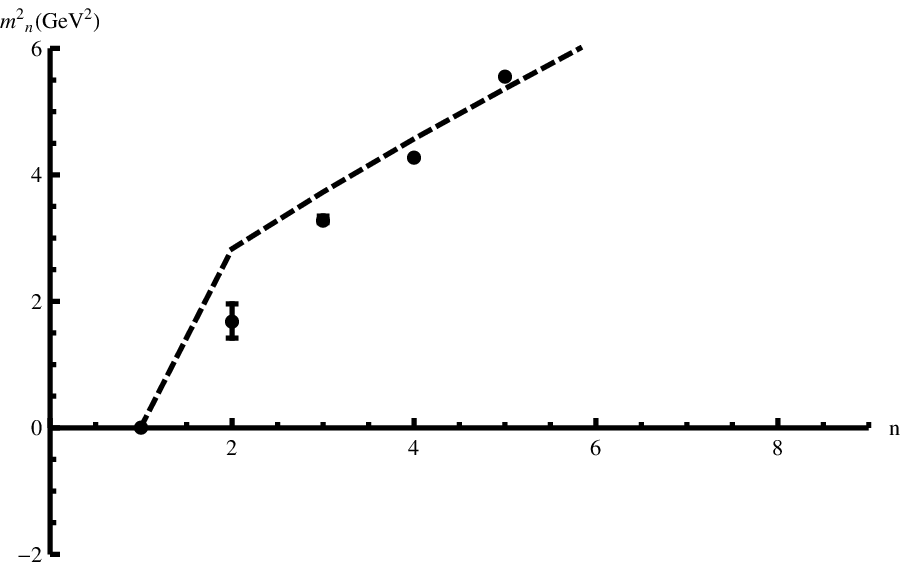} \vskip -0.05cm
\hskip 0.15 cm
\textbf{( Mod IIA ) } \hskip 6.5 cm \textbf{( Mod IIB )} \\
\end{center}
\caption[]{Pseudo-scalar spectra $m_{\pi,n}^2$ as functions of $n$  for Mod I and II
defined in Table \ref{parameters}. } \label{pscalarmassespic}
\end{figure}

\begin{figure}[h]
\begin{center}
\epsfxsize=10.0 cm \epsfysize=7.5 cm \epsfbox{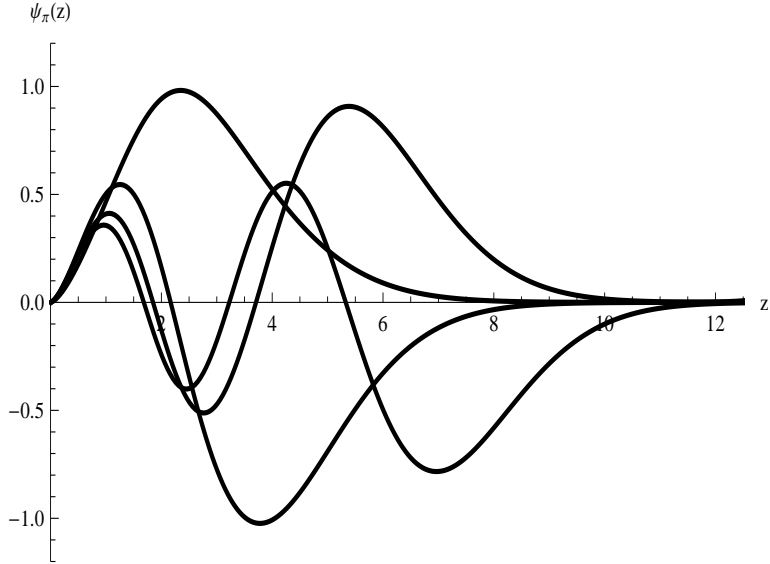}
\end{center}
\caption{The pseudoscalar wave function $\psi_{\pi,n}(z)$ as function of $z$ for Mod IB. }
\label{piwavepic}
\end{figure}

The equations of motion for the pesudoscalar $\pi$ coupled with $\varphi$ take the form of
\begin{eqnarray}
 -e^{-(3A_s-\Phi)}\partial_z(e^{3A_s-\Phi}\chi^2
 \partial_z)\pi+q^2\chi^2(\pi-\varphi)=&&0, \\
 -e^{-(A_s-\Phi)}\partial_z(e^{A_s-\Phi}
 \partial_z)\varphi-g_5^2\chi^2e^{2A_s}(\pi-\varphi)=&&0,
\end{eqnarray}
which can be written in the following form
\begin{eqnarray}
 -&&\pi_n''+V_{\pi,\varphi}\pi_n=m_n^2(\pi_n-e^{A_s}\chi\varphi_n), \nonumber\\
 -&&\varphi_n''+ V_{\varphi} \varphi_n=g_5^2 e^{A_s}\chi(\pi_n-e^{A_s}\chi\varphi_n),
\end{eqnarray}
with the effective schrodinger potentials
\begin{eqnarray}
V_{\pi,\varphi}&=& \frac{3A_s^{''}-\Phi^{''}+2\chi^{''}/\chi-2\chi^{'2}/\chi^2}{2}
 +\frac{(3A_s^{'}-\Phi^{'}+2\chi^{'}/\chi)^2}{4}, \nonumber \\
V_{\varphi} &=& \frac{A_s^{''}-\Phi^{''}}{2}+\frac{(A_s^{'}-\Phi^{'})^2}{4}.
\end{eqnarray}

With the two sets of parameters
given in Table \ref{parameters} for Mod I and Mod II, the pseudoscalar spectra $\pi$ are shown
in Table \ref{pscalarmasses} and Fig. \ref{pscalarmassespic}, and the corresponding
wave-functions are shown in Fig. \ref{piwavepic}.
It is observed that in our graviton-dilaton-scalar system, the lowest pseudoscalar state has
a mass around $140 {\rm MeV}$, which can be regarded as the Nambu-Goldstone bosons due
to the chiral symmetry breaking. The higher excitations behave a Regge line which agrees
well with experimental data.

\subsubsection{Vector sector}

In the vector sector, the terms of quadratic order in $V^\bot$ are
\begin{eqnarray}
 S_{V}^{(2)}=-\frac{N_f}{2g_5^2N_c L^3}\int d^5x
 e^{-\Phi}b_s^5\big(\partial_z V^\bot_\mu \partial^z V^{\bot\mu}
 +\partial_\mu V^\bot_\nu\partial^\mu V^{\bot\nu}\big),
\end{eqnarray}

\begin{table}
\begin{center}
\begin{tabular}{cccccccc}
\hline\hline
        n & $\rho$~exp.~(MeV)     & Mod~IA.~(MeV)         & Mod~IB.~(MeV)    &Mod~IIA.~(MeV)   &Mod~IIB.~(MeV) \\ \hline
        1 & $775.5 \pm 1$         &728                    & 771               &754             &797   \\
        2 & $1282 \pm 37$         &1135                   & 1143              &1134            &1140                         \\
        3 & $1465 \pm 25$         &1425                   & 1431              &1429            &1432\\
        4 & $1720 \pm 20$         &1665                   & 1670              &1668            &1672 \\
        5 & $1909 \pm 30$         &1874                   & 1878              &1876            &1880\\
        6 & $2149 \pm 17$         &2062                   & 2065              &2063            &2067\\
        7 & $2265 \pm 40$         &2234                   & 2237              &2235            &2238\\
\hline\hline
\end{tabular}
\caption{The experimental and predicted mass spectra for vector
mesons $\rho$.} \label{vectormasses}
\end{center}
\end{table}

\begin{figure}[h]
\begin{center}
\epsfxsize=6.5 cm \epsfysize=6.5 cm \epsfbox{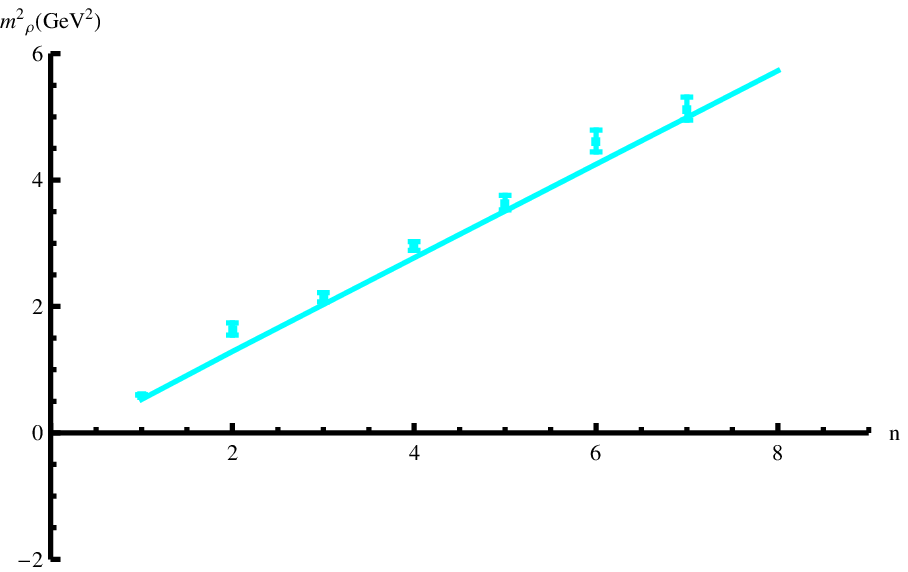} \hspace*{0.1cm}
\epsfxsize=6.5 cm \epsfysize=6.5 cm \epsfbox{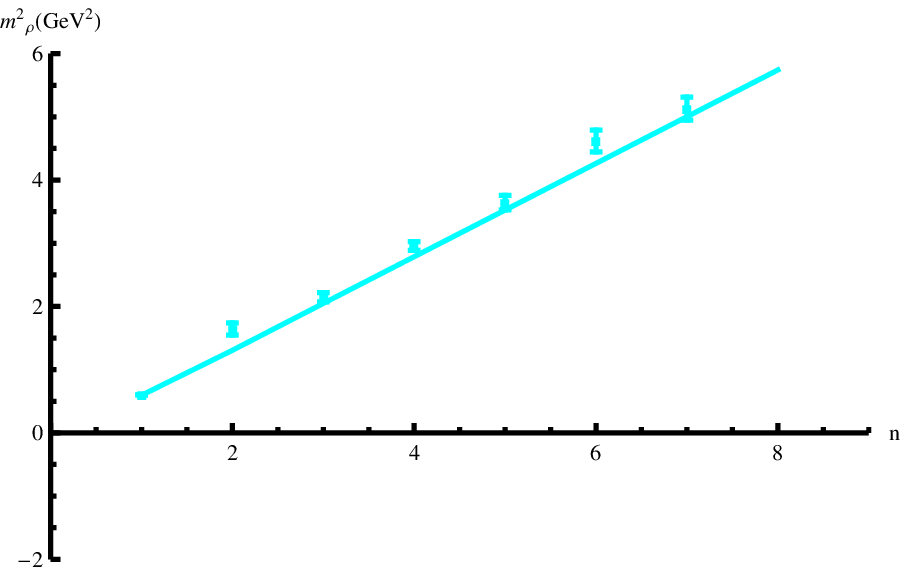} \vskip -0.05cm
\hskip 0.15 cm
\textbf{( Mod IA ) } \hskip 6.5 cm \textbf{( Mod IB )} \\
\epsfxsize=6.5 cm \epsfysize=6.5 cm \epsfbox{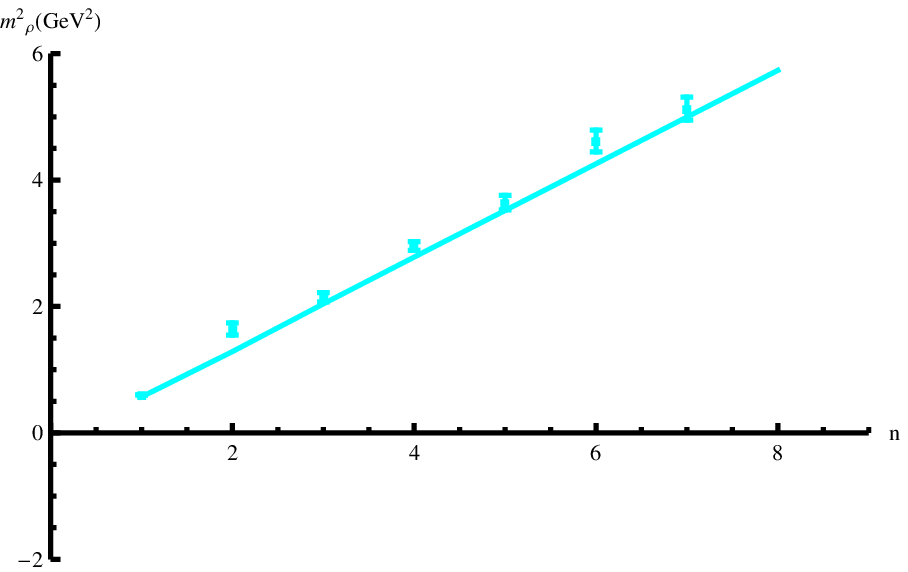} \hspace*{0.1cm}
\epsfxsize=6.5 cm \epsfysize=6.5 cm \epsfbox{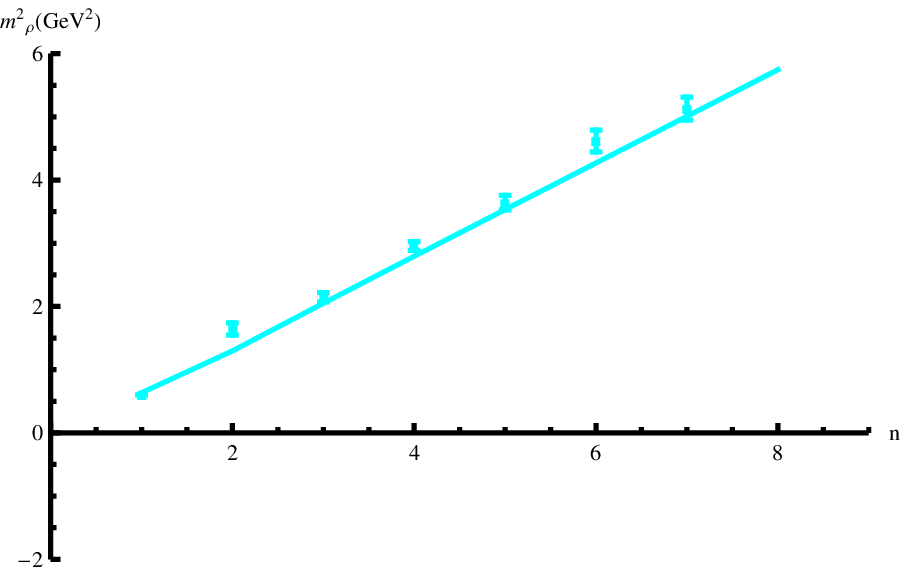} \vskip -0.05cm
\hskip 0.15 cm
\textbf{( Mod IIA ) } \hskip 6.5 cm \textbf{( Mod IIB )} \\
\end{center}
\caption[]{$m_{\rho,n}^2$ as functions of $n$  for Mod I and II defined in Table \ref{parameters}.} \label{vectormassespic}
\end{figure}

\begin{figure}[h]
\begin{center}
\epsfxsize=10.0 cm \epsfysize=7.5 cm \epsfbox{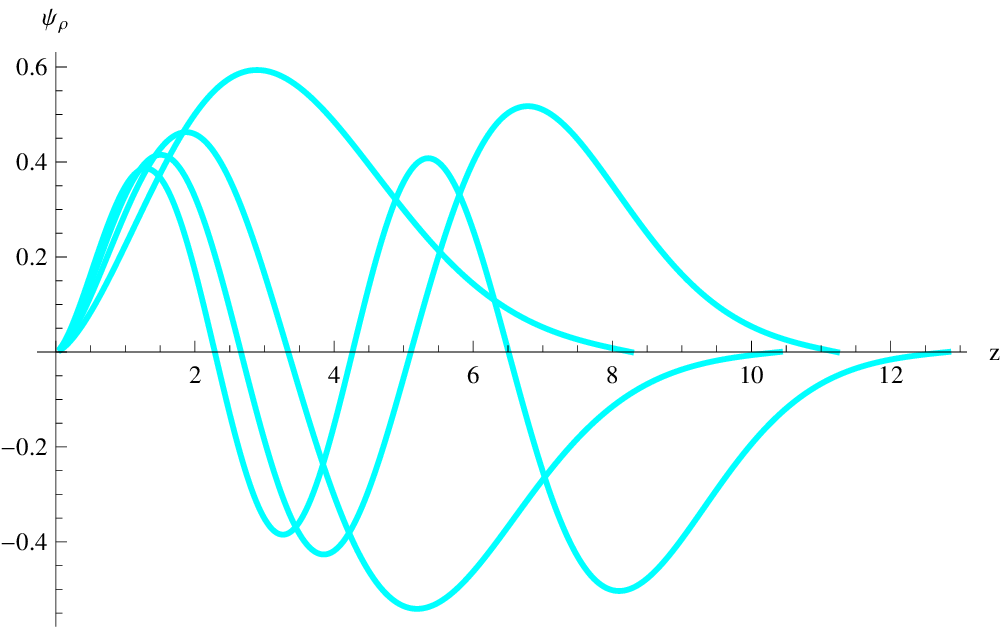}
\end{center}
\caption{Pseudoscalar wavefunction $\psi_{\rho,n}(z)$ as function
of $z$ for Mod IB. }
\label{rhowavepic}
\end{figure}

The equations of motion of the vector mesons take the form of
\begin{eqnarray}
-\rho_n^{''}+V_v\rho_n=m_n^2 \rho_n,
\end{eqnarray}
with the schrodinger like potential
\begin{equation}
V_v=\frac{A_s^{'}-\Phi^{'}}{2}+\frac{(A_s^{'}-\Phi^{'})^2}{4}.
\label{V_v}
\end{equation}

With the two sets of parameters
given in Table \ref{parameters} for Mod I and Mod II, vector spectra are shown
in Table \ref{vectormasses} and Fig. \ref{vectormassespic}, and the corresponding
wave-functions are shown in Fig. \ref{rhowavepic}.
It is observed that in our graviton-dilaton-scalar system, the lowest vector state has
a mass around $770 MeV$, and the higher excitations behave a Regge line which agrees well with
experimental data.

\subsubsection{Axial Vector Sector}
The terms of quadratic order in $A^\bot$ is
\begin{eqnarray}
 S_{A}^{(2)}=-\frac{N_f}{2g_5^2N_c L^3}\int d^5x
 e^{-\Phi}b_s^5\big(\partial_z A^\bot_\mu \partial^z A^{\bot\mu}
 +\partial_\mu A^\bot_\nu\partial^\mu A^{\bot\nu}+\frac{g_5^2\chi^2}{L^2} A^\bot_\mu A^{\bot\mu}\big).
\end{eqnarray}

\begin{table}
\begin{center}
\begin{tabular}{cccccccc}
\hline\hline
        n & $a_1$~Exp ~(MeV)        & Mod ~IA~(MeV)    & Mod~IB~(MeV)         & Mod ~IIA~(MeV)    & Mod~IIB~(MeV)  \\
\hline
        1 & $1230 \pm 40$           &1065              & 1316                 &1118               &1340\\
        2 & $1647 \pm 22$           &1562              & 1735                 &1625               &1753\\
        3 & $1930^{+30}_{-70}$      &1846              & 1969                 &1879               &1979\\
        4 & $2096 \pm 122$          &2058              & 2163                 &2083               &2168\\
        5 & $2270^{+55}_{-40}$      &2243              & 2336                 &2264               &2339\\
\hline\hline
\end{tabular}
\caption{The experimental and predicted mass spectra for axial
vector mesons $a_1$.} \label{avectormasses}
\end{center}
\end{table}

\begin{figure}[h]
\begin{center}
\epsfxsize=6.5 cm \epsfysize=6.5 cm \epsfbox{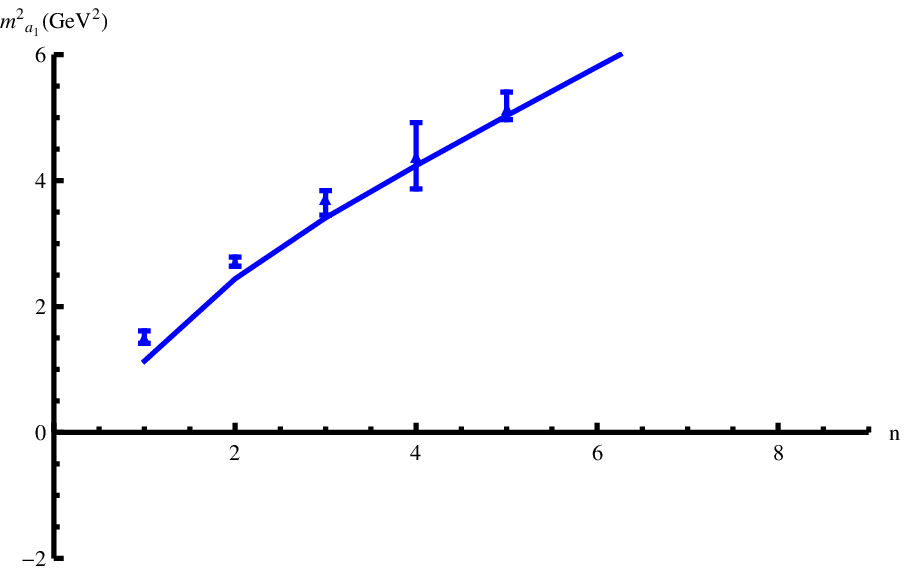} \hspace*{0.1cm}
\epsfxsize=6.5 cm \epsfysize=6.5 cm \epsfbox{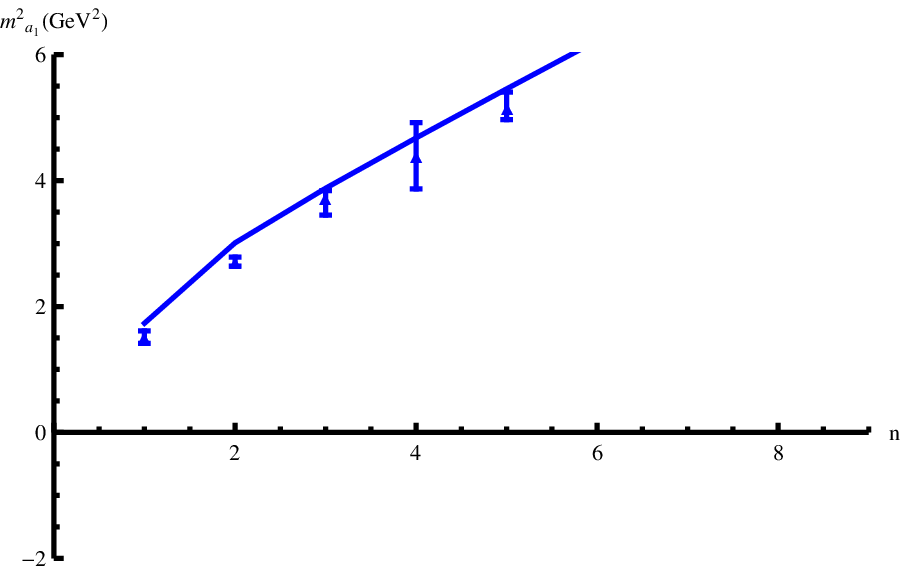} \vskip -0.05cm
\hskip 0.15 cm
\textbf{( Mod IA ) } \hskip 6.5 cm \textbf{( Mod IB )} \\
\epsfxsize=6.5 cm \epsfysize=6.5 cm \epsfbox{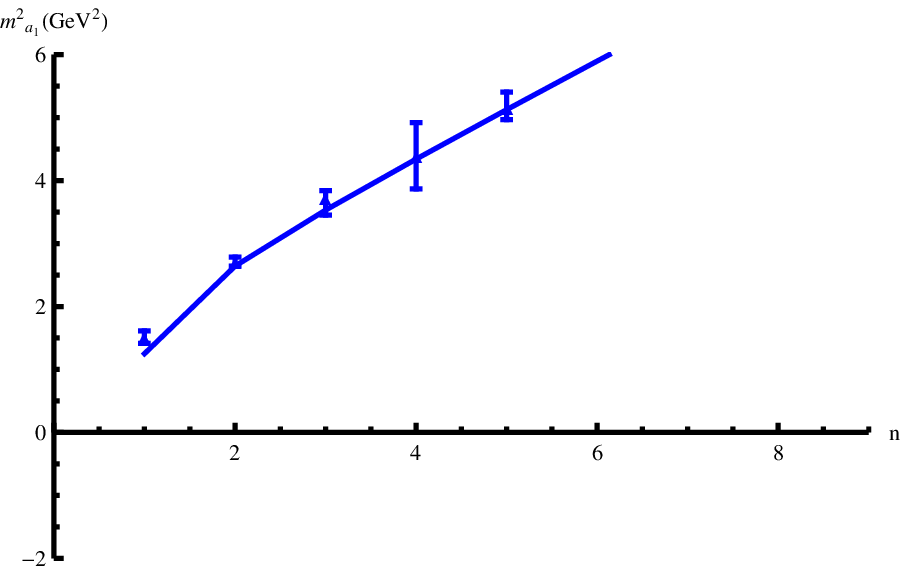} \hspace*{0.1cm}
\epsfxsize=6.5 cm \epsfysize=6.5 cm \epsfbox{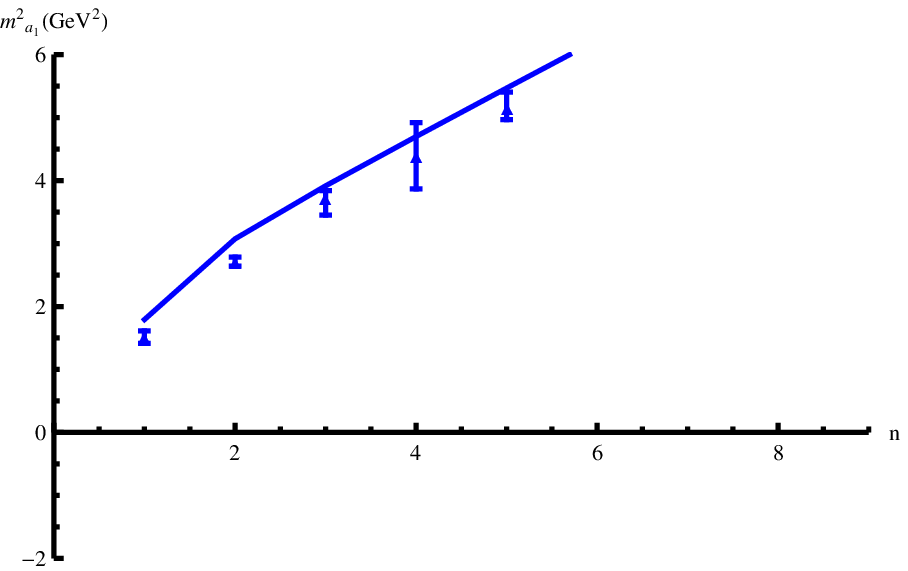} \vskip -0.05cm
\hskip 0.15 cm
\textbf{( Mod IIA ) } \hskip 6.5 cm \textbf{( Mod IIB )} \\
\end{center}
\caption[]{$m_{a_1,n}^2$ as functions of $n$ for Mod I and II defined in
Table \ref{parameters}. } \label{avectormassespic}
\end{figure}

\begin{figure}[h]
\begin{center}
\epsfxsize=10.0 cm \epsfysize=7.5 cm \epsfbox{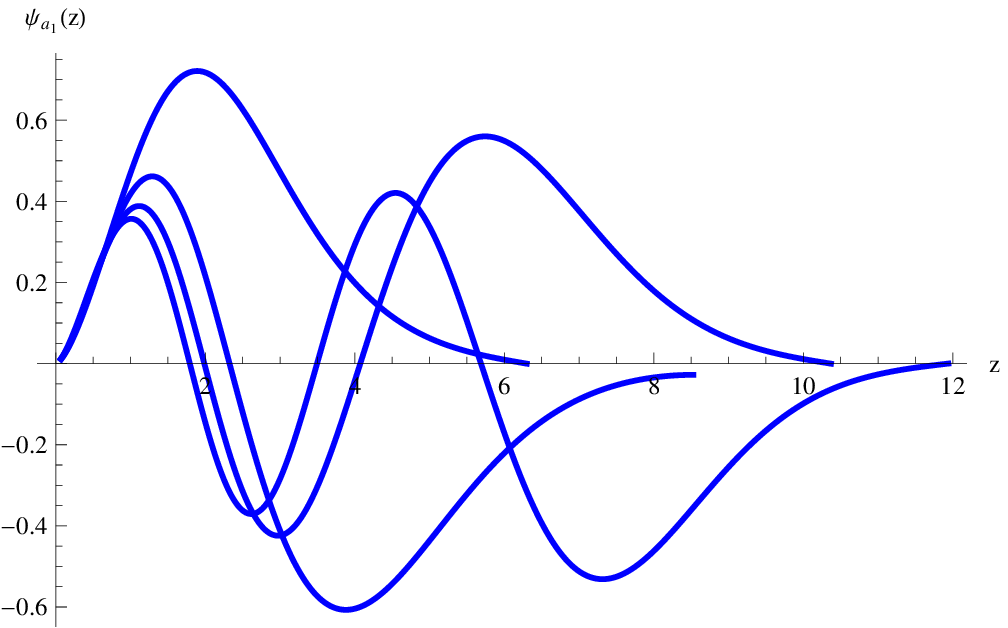}
\end{center}
\caption{Axial-vector meson wave-function $\psi_{a_1,n}(z)$ as
function of $z$ for Mod IB. }
\label{a1wavepic}
\end{figure}

The equations of motion of the axial-vector mesons take the form of:
\begin{eqnarray}
-a_n^{''}+V_a a_n&=& m_n^2 a_n,
\end{eqnarray}
with the schrodinger potential for the axial vector as
\begin{equation}
V_a=\frac{A_s^{'}-\Phi^{'}}{2}+\frac{(A_s^{'}-\Phi^{'})^2}{4}+g_5^2 e^{2A_s}\chi^{2}.
\label{V_a}
\end{equation}

As we can see that the difference between the schrodinger potentials for the axial vector
Eq.(\ref{V_a}) and vector Eq.(\ref{V_v}) is only the extra term $g_5^2 e^{2A_s} \chi^{2}$
in Eq.(\ref{V_a}). In the KKSS model, $g_5^2 e^{2A_s} \chi^{2}\rightarrow 0$ when
$z\rightarrow\infty$, therefore there is no splitting between vector and axial vector.
Here in the graviton-dilaton-scalar system, $g_5^2 e^{2A_s} \chi^{2}\rightarrow constant$
when $z\rightarrow\infty$, which naturally induces the separation of the chiral partners.

With the two sets of parameters
given in Table \ref{parameters} for Mod I and Mod II, axial vector spectra are shown
in Table \ref{avectormasses} and Fig. \ref{avectormassespic}, and the corresponding
wave-functions are shown in Fig. \ref{a1wavepic}.
It is observed that in our graviton-dilaton-scalar system, the lowest axial vector state
has a mass around the experimental value $1230 MeV$, and
the higher excitations behave a Regge line which agrees well with
experimental data.

\subsection{Short summary}

\begin{figure}[h]
\begin{center}
\epsfxsize=6.5 cm \epsfysize=6.5 cm \epsfbox{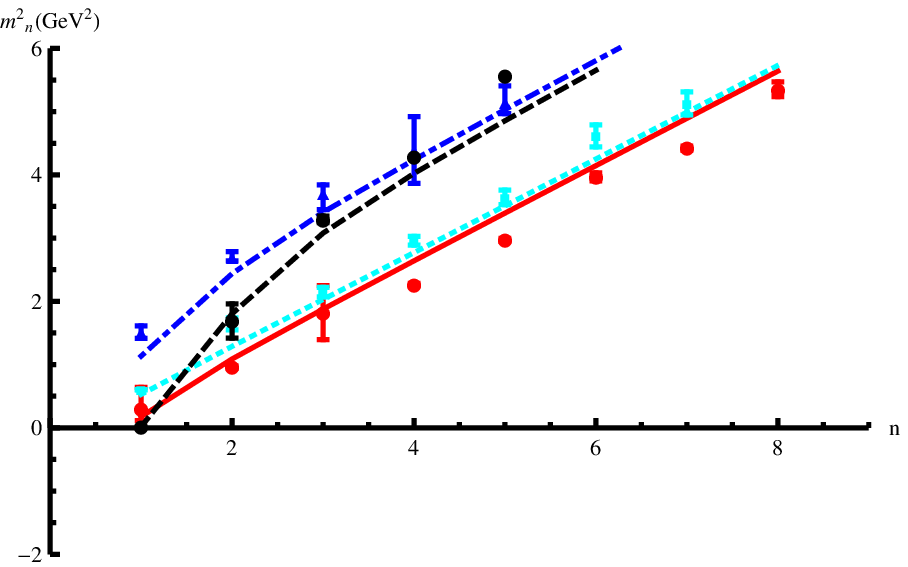} \hspace*{0.1cm}
\epsfxsize=6.5 cm \epsfysize=6.5 cm \epsfbox{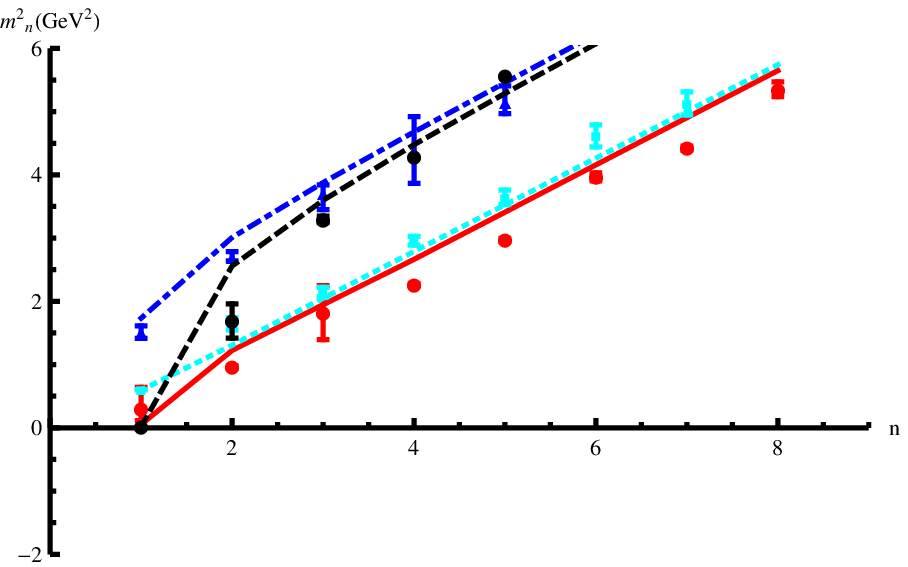} \vskip -0.05cm
\hskip 0.15 cm
\textbf{( Mod IA ) } \hskip 6.5 cm \textbf{( Mod IB )} \\
\epsfxsize=6.5 cm \epsfysize=6.5 cm \epsfbox{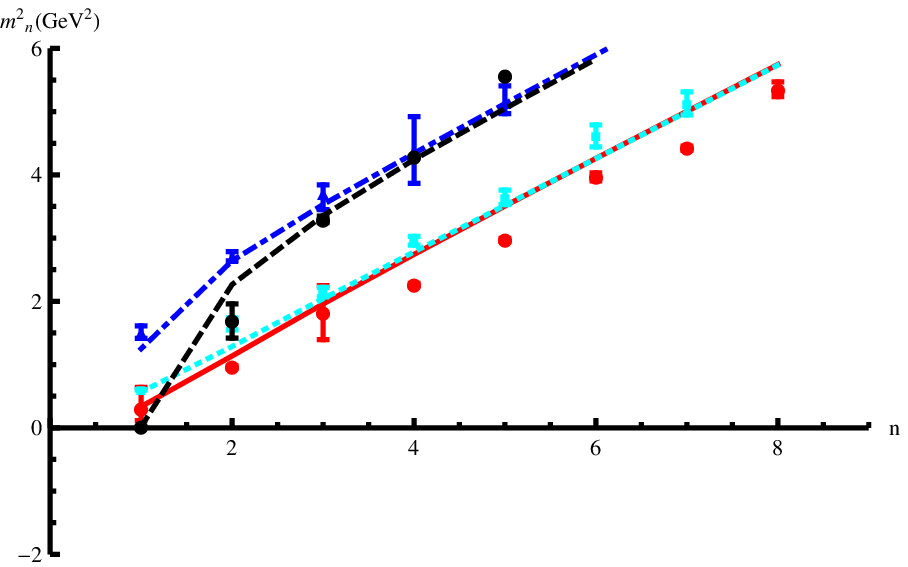} \hspace*{0.1cm}
\epsfxsize=6.5 cm \epsfysize=6.5 cm \epsfbox{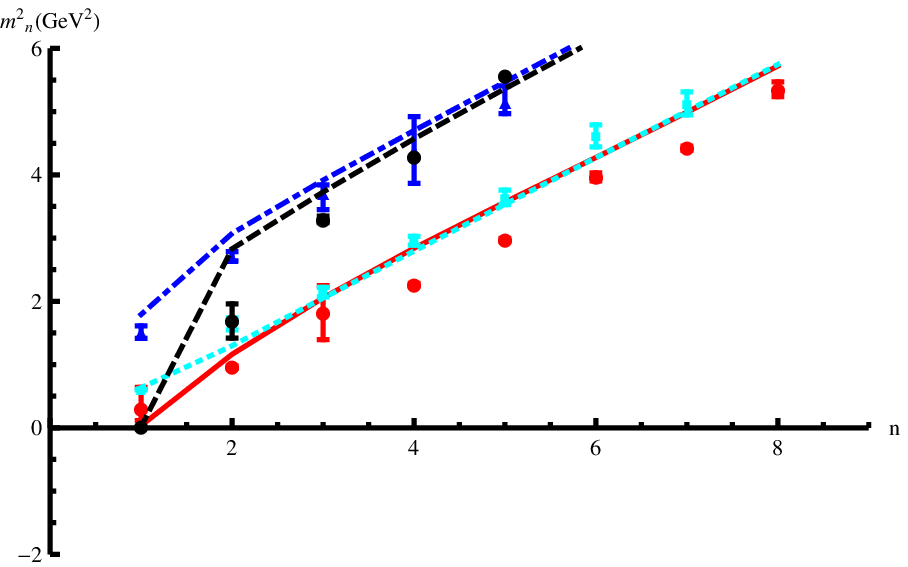} \vskip -0.05cm
\hskip 0.15 cm
\textbf{( Mod IIA ) } \hskip 6.5 cm \textbf{( Mod IIB )} \\
\end{center}
\caption{All meson spectra in Mod I and Mod II with two sets
of parameters in Table \ref{parameters} comparing with experimental data.}
\label{allmassespic}
\end{figure}

We summarize the meson spectra for scalar, pseudoscalar, vector and axial-vector
in Fig. \ref{allmassespic} for the two sets of parameters. It is found that for both
cases, the results are in very well agreement with experimental data. The ground
state has the order of $m_{\pi}<m_{f_0}<m_{\rho}<m_{a_1}$, and the Regge slopes
for scalar, pseudoscalar, vector and axial-vector meson at high excitations take
the same value of $4\mu_G^2$.

\section{Decay constants, pion form factor and vector couplings}
\label{sec-formfactors}

So far, within the above holography model we have studied one aspect of the static hadronic properties, the resonance masses. To further confirm this model, we have to check that whether it can reproduce reasonable behavior of other static and dynamic properties of hadronic physics, such as decay constants, vector couplings and form factors etc.

As in Ref.\cite{EKSS2005} (for more details see Ref. \cite{Hong:2004sa}), by studying the current-current correlation function and rewriting it as a summation over the normalizable wave functions, we can extract the decay constants $f_\pi, F_{\rho_n},F_{a_1,n}$ as following,
\begin{eqnarray}
f_\pi^2&&=-\frac{N_f}{g_5^2N_c}e^{A_s-\Phi}\partial_z A(0,z)|_{z\rightarrow0}, \\
F_{\rho_n}^2&&=\frac{N_f}{g_5^2N_c}(e^{A_s-\Phi}\partial_z V_n(z)|_{z\rightarrow0})^2, \\
F_{a_1,n}^2&&=\frac{N_f}{g_5^2N_c}(e^{A_s-\Phi}\partial_z A_n(z)|_{z\rightarrow0})^2.
\end{eqnarray}
Where $A(0,z),V_n(z), A_n(z)$ is the solution of equations
\begin{eqnarray}
 (-e^{-(A_s-\Phi)}\partial_z(e^{A_s-\Phi}\partial_z
 )+g5^2 e^{2A_s}\chi^2)A(0,z)=0,\\
(-e^{-(A_s-\Phi)}\partial_z(e^{A_s-\Phi}\partial_z
 )-m_{\rho,n}^2)V_n(z)=0,\\
(-e^{-(A_s-\Phi)}\partial_z(e^{A_s-\Phi}\partial_z
 )+g_5^2 e^{2 A_s}\chi^2-m_{a_1,n}^2)A_n(z)=0,
\end{eqnarray}
with the boundary condition$A(0,0)=1,\partial_z A(0,\infty)=0$, $V_n(0)=0,\partial_zV_n(\infty)=0$,$A_n(0)=0,\partial_zA_n(\infty)=0$ and normalized as  $\int dz e^{A_s-\Phi} V_m V_n=\int dz e^{A_s-\Phi} A_m A_n=\delta_{mn}.$

We can also extract the pion form factor from the three point correlator as \cite{Grigoryan:2007wn,Grigoryan:2007vg,Kwee:2007dd,Kwee:2007nq}
\begin{eqnarray}
  f_\pi^2 F_\pi(Q^2)=\frac{N_f}{g_5^2N_c}\int dz e^{A_s-\Phi} V(q^2,z) \big\{ (\partial_z\varphi)^2+g_5^2\chi^2 e^{2A_s} (\pi-\varphi)^2\big\},
\end{eqnarray}
where$Q^2=-q^2$, and $V(q^2,z),\pi(z),\varphi(z)$ is the solution of
\begin{eqnarray}
(-e^{-(A_s-\Phi)}\partial_z(e^{A_s-\Phi}\partial_z
 )+q^2)V(q^2,z)=0,\\
  -e^{-(3A_s-\phi)}\partial_z(e^{3A_s-\phi}\chi^2
 \partial_z)\pi-m_{\pi,n}^2\chi^2(\pi-\varphi)=0, \\
 -e^{-(A_s-\phi)}\partial_z(e^{A_s-\phi}
 \partial_z)\varphi-g_5^2\chi^2e^{2A_s}(\pi-\varphi)=0,
\end{eqnarray}
with the boundary condition $V(q^2,0)=1,\partial_zV(q^2,\infty)=0, \pi(0)=0,\partial_z\pi(\infty)=0,\varphi(0)=0,\varphi(\infty)=0$ and normalized as
\begin{eqnarray}
\frac{N_f}{g_5^2 N_c f_\pi^2}\int dz e^{A_s-\Phi} \big\{ (\partial_z\varphi)^2+g_5^2\chi^2 e^{2A_s} (\pi-\varphi)^2\big\}=1.
\end{eqnarray}

To make sure that $F_\pi(0)=1$ or equivalently we can write
\begin{eqnarray}
 F_\pi(Q^2)=\frac{\int dz e^{A_s-\Phi} V(q^2,z) \big\{ (\partial_z\varphi)^2+g_5^2\chi^2 e^{2A_s} (\pi-\varphi)^2\big\}}{\int dz e^{A_s-\Phi} \big\{ (\partial_z\varphi)^2+g_5^2\chi^2 e^{2A_s} (\pi-\varphi)^2\big\}}.
\end{eqnarray}

Decomposing $F_\pi$ as in \cite{Grigoryan:2007wn,Grigoryan:2007vg,Kwee:2007dd,Kwee:2007nq}, we reach
\begin{eqnarray}
F_\pi(Q^2)=\sum_n \frac{F_{\rho_n} g_{n\pi \pi}}{Q^2+m_n^2},
\end{eqnarray}
with
\begin{eqnarray}
g_{n\pi\pi}=g_5 \frac{\int dz e^{A_s-\Phi} V_n \big\{ (\partial_z\varphi)^2+g_5^2\chi^2 e^{2A_s} (\pi-\varphi)^2\big\}}{\int dz e^{A_s-\Phi} \big\{ (\partial_z\varphi)^2+g_5^2\chi^2 e^{2A_s} (\pi-\varphi)^2\big\}}.
\end{eqnarray}
We would denote $g_{\rho\pi\pi}\equiv g_{0\pi\pi}$, i.e. putting the $\rho$ meson ground state wave function $V_\rho\equiv V_0$ in the above equation.

The numerical results for the decay constant are shown in Table \ref{decay-form-vcoupling}
and are compared with other models in Table \ref{decay-form-vcoupling-compare}.

\begin{table}
\begin{center}
\begin{tabular}{cccccccc}
\hline\hline
     & ~exp.~(MeV) & Mod~IA & Mod~IB &Mod~IIA &Mod~IIB    \\  \hline
     $f_\pi$   & $92.4\pm0.35$   & 59.3   & 83.6  &65.7    &87.4\\
        $F_\rho^{1/2}$  & $346.2 \pm 1.4$   & 270  &282   &290    &299\\
       $F_{a_1}^{1/2}$  & $433 \pm 13$   & 379   &452    &411     &474\\
      $g_{\rho\pi\pi}$  & $6.03 \pm 0.07$  & 4.63  &3.14   &4.41  &3.17\\
\hline\hline
\end{tabular}
\caption{Decay constant in Mod I and Mod II with two sets of parameters in
Table \ref{parameters}, and the unit is in ${\rm MeV}$.}
\label{decay-form-vcoupling}
\end{center}
\end{table}

\begin{table}
\begin{center}
\begin{tabular}{cccccccc}
\hline\hline
 & ~exp.~(MeV) & Mod~IB & KKSS  & hard-wall  & mod-soft  &SWXY & GKK \\ \hline
 $f_\pi$ & $92.4\pm0.35$ & 83.6 & 87.0 &92.1 &88.0 &92.4  &92.4\\
 $F_\rho^{1/2}$ & $346.2 \pm 1.4$    & 282   &261  &329   &325   &---    &---\\
 $F_{a_1}^{1/2}$ & $433 \pm 13$  & 452 &558  &463  &474   &---   &---\\
 $g_{\rho\pi\pi}$ & $6.03 \pm 0.07$  & 3.14  &3.33 &4.48  &4.63   &3.51   &2.89\\
\hline\hline
\end{tabular}
\caption{Table for decay constants and couplings from other models, the results in
hard-wall,KKSS model stands, mod-soft are taken from \cite{Kwee:2007dd,Kwee:2007nq}.}
\label{decay-form-vcoupling-compare}
\end{center}
\end{table}

We can see that under our parametrization, both in Mod I and Mod II a larger $\sigma$ would give a larger $f_\pi$, and with parameters in set B, the deviations of $f_\pi$ from the experimental data are within $10\%$ while in set A the smaller values of $\sigma$ make $f_\pi$ $40\%$ smaller than the experimental data. The $a_1$ decay constants $F_{a_1}^{1/2}$ prediction is closer to the experimental data than other models,
and both the prediction of the pion form factor
are better than the original soft-wall model in Ref.\cite{Kwee:2007dd,Kwee:2007nq},
while $F_\rho^{1/2},g_{\rho\pi\pi}$ are a little too small.
\begin{figure}[h]
\begin{center}
\epsfxsize=6.5 cm \epsfysize=6.5 cm \epsfbox{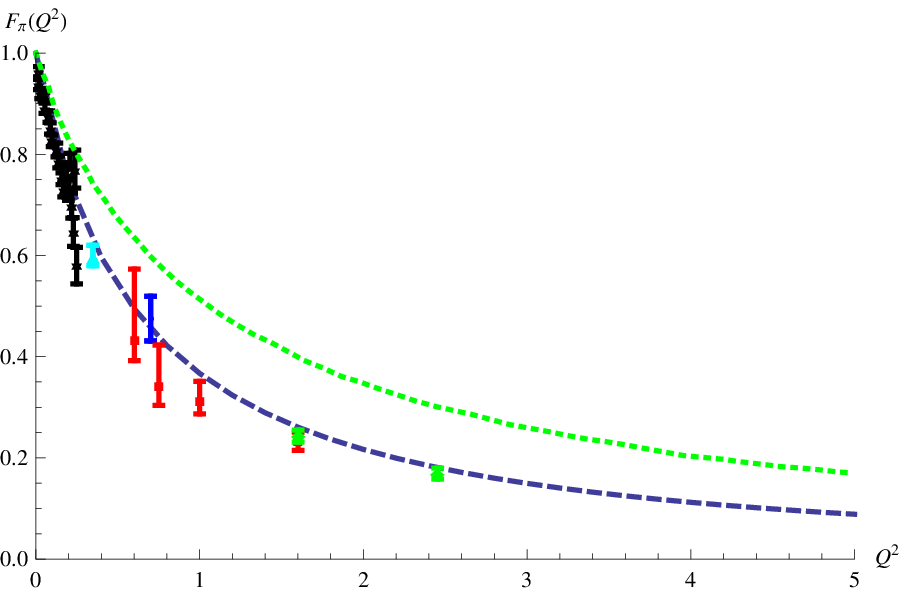} \hspace*{0.1cm}
\epsfxsize=6.5 cm \epsfysize=6.5 cm \epsfbox{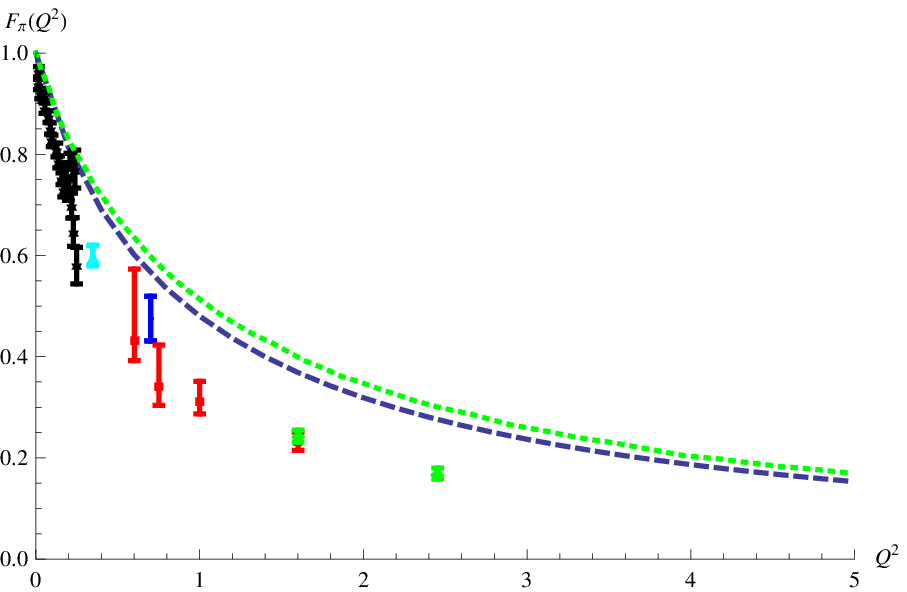} \vskip -0.05cm
\hskip 0.15 cm
\textbf{( Mod IA ) } \hskip 6.5 cm \textbf{( Mod IB )} \\
\epsfxsize=6.5 cm \epsfysize=6.5 cm \epsfbox{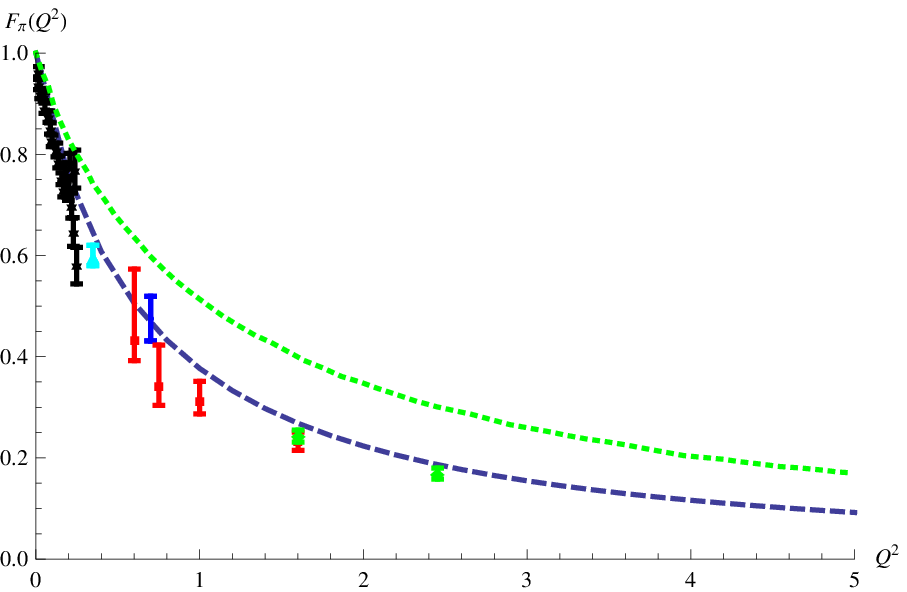} \hspace*{0.1cm}
\epsfxsize=6.5 cm \epsfysize=6.5 cm \epsfbox{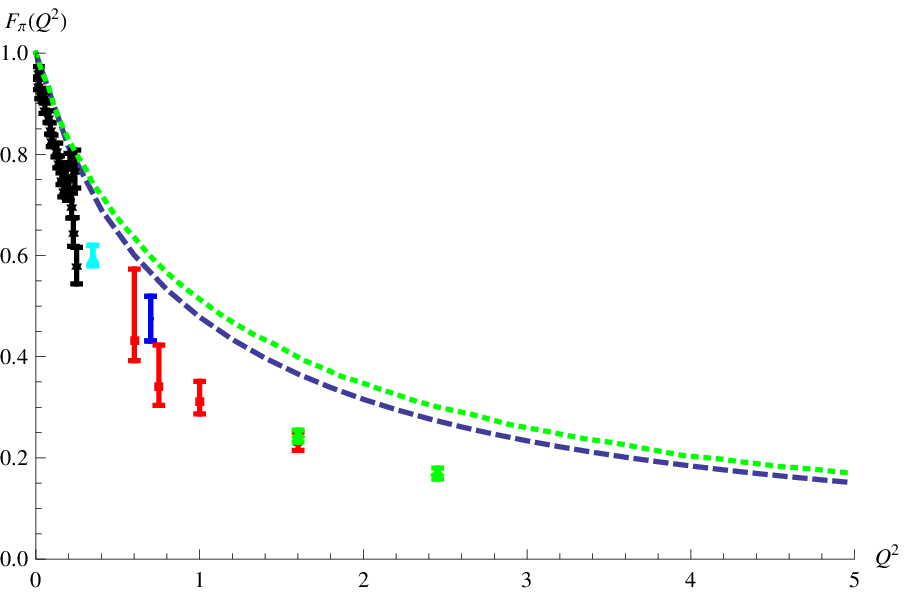} \vskip -0.05cm
\hskip 0.15 cm
\textbf{( Mod IIA ) } \hskip 6.5 cm \textbf{( Mod IIB )} \\
\end{center}
\caption[]{$F_\pi(Q^2)$ as function of $Q^2$ for Mod I and II defined in Table \ref{parameters}
and compared with experimental data. The blue dashed lines are the prediction in our model, and the green dotted line is the original soft-wall model results taken from Ref.\cite{Kwee:2007dd,Kwee:2007nq}.} \label{formfactors}
\end{figure}

The pion form factor is shown in Fig. \ref{formfactors}, it is found that with
parameters set A used for Mod I and Mod II with a smaller chiral condensate,
the produced pion form factor matches the experimental data much better, however,
the produced pion decay constant is much smaller than experimental data as shown
in Table \ref{decay-form-vcoupling}. With parameters in set B for both Mod I and Mod II
corresponding to a larger chiral condensate, one can produce better result for pion decay constant, but the results on pion form factor are worse.

\section{Discussion and summary}
\label{sec-summary}

In this work, we construct a quenched dynamical holographic QCD (hQCD) model
in the graviton-dilaton framework for the pure gluon system, and develop a dynamical
hQCD model for the two flavor system in the graviton-dilaton-scalar framework
by adding light flavors on the gluodynamical background.
Two forms of dilaton background field $\Phi=\mu_G^2z^2$ and $\Phi=\mu_G^2z^2\tanh(\mu_{G^2}^4z^2/\mu_G^2)$ have been considered in this work.
In both cases, the quadratic correction to dilaton background field at IR encodes
important non-perturbative gluodynamics and naturally induces a deformed warp factor
of the metric. With the pure quadratic dilaton background field $\Phi=\mu_G^2z^2$,
the dynamical holographic model can be regarded as a selfconsistent
KKSS model, where the metric structure is not AdS$_5$ anymore but automatically deformed
at IR. However, with quadratic dilaton background field, one may encounter
the gauge invariant problem for the dimension-2 gluon operator. To avoid the gauge
non-invariant problem and to meet the requirement of gauge/gravity duality,
we propose the dilaton field with quartic form at UV and quadratic form at IR as in
$\Phi=\mu_G^2z^2\tanh(\mu_{G^2}^4z^2/\mu_G^2)$.

In the quenched dynamical model, without introducing extra parameters but just
self-consistently solving the deformed metric induced by the dilaton background field,
we find that the scalar glueball spectra is in very well agreement with lattice data,
while the soft-wall model with ${\rm AdS}_5$ metric cannot accommodate both the ground state
and the Regge slope for the scalar glueball spectra. We also give a necessary condition
for the existence of linear quark potential from the metric structure, and we show
that in the graviton-dilaton framework, a negative quadratic dilaton background
field cannot produce the linear quark potential.

For two flavor system in the graviton-dilaton-scalar framework, the deformed metric
is self-consistently solved by considering both the chiral condensate and non-perturbative
gluodynamics in the vacuum, which are responsible for the chiral symmetry breaking and
linear confinement, respectively. It is found that the mixing
between the chiral condensate and gluon condensate is important in the dynamical hQCD
model to produce the correct light flavor meson spectra.

The pion form factor and the vector couplings are also investigated in the dynamical
hQCD model. It is found that with smaller chiral condensate,
the produced pion form factor matches the experimental data much better, however,
the produced pion decay constant is much smaller than experimental data. With
larger chiral condensate, one can produce better result for pion decay constant,
but the result on pion form factor is worse.

In summary, we have offered a systematic framework to describe the
non-perturbative gluodynamics and chiral dynamics. The input in our model is basically
the non-perturbative gluodynamics represented by $\mu_G^2$, the chiral condensate $\sigma$,
and a current quark mass $m_q$, which are the same as in the soft-wall model. Just solve the deformed warp factor self-consistently, one can produce the glueball spectra, the linear
heavy quark potential as well as light flavor meson spectra in very well agreement with
lattice and experimental data.

\vskip 0.5cm
{\bf Acknowledgement}
\vskip 0.2cm

This work is supported by the NSFC under Grant
Nos. 11175251 and 11275213, DFG and NSFC (CRC 110),
CAS key project KJCX2-EW-N01, K.C.Wong Education Foundation, and
Youth Innovation Promotion Association of CAS.

\end{document}